\documentclass[sigconf]{acmart}

\AtBeginDocument{%
  \providecommand\BibTeX{{%
    \normalfont B\kern-0.5em{\scshape i\kern-0.25em b}\kern-0.8em\TeX}}}

\usepackage{subcaption}
\usepackage{multirow}
\usepackage{algorithmic}
\usepackage{algorithm}
\usepackage{todonotes}
\usepackage{capt-of}
\usepackage{pbox}
\usepackage{tikz}
\def\checkmark{\tikz\fill[scale=0.4](0,.35) -- (.25,0) -- (1,.7) -- (.25,.15) -- cycle;} 

\begin{document}
\pagestyle{plain}

\title{On Practicality of Using ARM TrustZone Trusted Execution Environment for Securing Programmable Logic Controllers}



\author{Zhiang Li}
\email{zhiangli@u.nus.edu}
\affiliation{%
  \institution{National University of Singapore}
  \country{Singapore}
}

\author{Daisuke Mashima, Wen Shei Ong, Ertem Esiner}
\email{{daisuke.m,wenshei.ong,e.esiner}@iarcs-create.edu.sg}
\affiliation{%
  \institution{Illinois Advanced Research Center at Singapore}
  \country{Singapore}
}

\author{Zbigniew Kalbarczyk}
\email{kalbarcz@illinois.edu} 
\affiliation{%
  \institution{University of Illinois Urbana-Champaign}
  \country{USA}
}

\author{Ee-Chien Chang}
\email{changec@comp.nus.edu.sg}
\affiliation{%
  \institution{National University of Singapore}
  \country{Singapore}
}

\renewcommand{\shortauthors}{Li, et al.}

\begin{abstract}
Programmable logic controllers (PLCs) are crucial devices for implementing automated control in various industrial control systems (ICS), such as smart power grids, water treatment systems, manufacturing, and transportation systems. Owing to their importance, PLCs are often the target of cyber attackers that are aiming at disrupting the operation of ICS, including the nation's critical infrastructure, by compromising the integrity of control logic execution. While a wide range of cybersecurity solutions for ICS have been proposed, they cannot counter strong adversaries with a foothold on the PLC devices, which could manipulate memory, I/O interface, or PLC logic itself. These days, many ICS devices in the market, including PLCs, run on ARM-based processors, and there is a promising security technology called ARM TrustZone, to offer a Trusted Execution Environment (TEE) on embedded devices. Envisioning that such a hardware-assisted security feature becomes available for ICS devices in the near future, this paper investigates the application of the ARM TrustZone TEE technology for enhancing the security of PLC. Our aim is to evaluate the feasibility and practicality of the TEE-based PLCs through the proof-of-concept design and implementation using open-source software such as OP-TEE and OpenPLC. Our evaluation assesses the performance and resource consumption in real-world ICS configurations, and based on the results, we discuss bottlenecks in the OP-TEE secure OS towards a large-scale ICS and desired changes for its application on ICS devices. Our implementation is made available to public for further study and research.  
\end{abstract}


\maketitle

\thispagestyle{plain}

\section{Introduction}

In the constantly evolving landscape of Industrial control systems (ICS), which are cyber-physical systems widely adopted in modernized critical infrastructures such as power grids, water treatment/distribution systems, urban transportation systems, and so forth, the intersection of cybersecurity and physical system integrity has emerged as a critical concern. With the advent of digital technologies, ICS have started adopting Information and Communication Technology (ICT) solutions, e.g., networking over Ethernet and protocols like TCP/IP, to enjoy operational efficiency and interconnectivity. However, the connectivity with other systems and public networks implies that ICS face the same threats as enterprise IT systems. An attacker may penetrate the control system (also called operational technology, or OT, system) via an enterprise IT system that is connected to the Internet, often through VPN (virtual private network) interfaces for remote control and maintenance.

Programmable logic controllers (PLC) are crucial components in many kinds of ICS. Typically, PLCs consist of a closed control loop for automated control in ICS. PLCs retrieve measurements from various sensors, execute control logic, and then send out control commands to actuators when certain conditions defined in the logic are met. A series of these tasks are also called {\it scan cycle}.
Given these important roles that PLCs play, the integrity of their functionality is essential to the reliable and trustworthy operation of ICS. For instance, if the logic is maliciously modified, necessary protection actions for preventing damage to physical plants may not be taken. A similar consequence would happen if the sensor measurements reported to PLCs and outgoing control commands are forged or tampered with. Because of the criticality, PLCs are often attractive targets of cyber attackers. 
In this paper, we focus on the integrity of automated control executed by PLCs, including control logic and input/output for it, which is crucial to the reliability of ICS. 

In the literature, there is substantial evidence of regular attacks against ICS, including PLCs housed within these systems~\cite{ics-cert}. Prominent real-world examples such as Stuxnet~\cite{stuxnet}, 
and Havex~\cite{havex} have caused significant losses and impacts, posing severe threats to both the environment and daily life. Further substantiating these concerns, recent reports have identified multiple critical vulnerabilities in WAGO PLCs that could potentially grant attackers complete control over these systems~\cite{wago-vuln}.
Being aware of such risks, ICS/OT operators have employed advanced countermeasures against cyber threats in their systems. Industrial firewalls are set up to enforce network flow control policies; antivirus software is installed in control centers and engineering stations to block phishing e-mail and malware; corporate IT network uses secure network protocols to establish trusted communication. 
However, these are still not sufficient to counter sophisticated attacks. For instance, 
Triton~\cite{triton}, which caused operational disruption at a Saudi Arabian oil and gas facility, allowed attackers to manipulate arbitrary parameters of Schneider Electric’s Safety Instrumented System (SIS), which was reported in 2017 when antivirus software and firewalls were already quite common in ICS.

In recent years, advanced security solutions dedicated to ICS or cyber-physical systems have been proposed to detect sophisticated attacks, such as intrusion detection systems~\cite{tan2022cotoru}, message authentication~\cite{esiner2022lomos,tefek2022caching}, and so forth. 
Unfortunately, even with these solutions, assurance of the integrity of the control loop implemented by PLCs under a strong attacker model is still an open challenge.
In particular, once an attacker gets the capability of remote code execution or even a foothold on the PLC devices, these measures may be circumvented (e.g., by compromising the I/O buffer on the device). Thus, it is essential to deploy a robust security mechanism on PLC devices themselves to ensure security even against such attackers.  
%
%


In order to further harden the integrity of PLCs' control loop against attacks, we resort to security mechanisms with hardware support. 
Nowadays, commodity PLC products have started adopting ARM microprocessors~\cite{arm-plc}. For instance, PLC products from ABB, Mitsubishi, and WAGO run on ARM Cortex A8 or A9 processors~\cite{abbplc, melplc, wagoplc}.
Given the support from prominent market leaders, there is an anticipated increase in the adoption of ARM-based processors.
A promising security technology led by ARM is called ARM TrustZone~\cite{trustzone}, which Cortex A8/A9 processors technically support~\cite{cortexa8}. TrustZone is a hardware-level security extension for ARM SoCs to implement Trusted Execution Environment (TEE)~\cite{tee}. TEE offers a ``secure world'' for sensitive module execution and data storage, which is isolated from the Rich Execution Environment (REE, or also called the ``normal world''). TEE and TrustZone technology have been increasingly used in mobile phones and tablets to protect sensitive data and modules over the past decades, but their usage on ICS devices is still in an early stage. 
Envisioning that ARM TrustZone will be adopted on more ICS devices, including PLCs, in the near future, 
in this work, we explore the feasibility and practicality of using ARM TrustZone for securing the control loop implemented by PLCs through the proof-of-concept implementation, called TEE-PLC. 
{\color{black}To our knowledge, our work is the first to 
utilize ARM TrustZone TEE for securing PLC's scan cycle integrity and implement a proof-of-concept for 
studying the practicality and limitation.}
ARM TrustZone is designed primarily for integrity and confidentiality, but not for real-time processing, including interactions with networked devices, and thus this paper will highlight the design considerations and recommendations for future TEE and secure OS implementation suitable for PLCs.  



Our contribution includes the following: 
\begin{itemize}
    \item Summarize threat models against the integrity of PLC control logic and enumerate attack vectors and discuss how TEE technology can mitigate them.
    \item Design {\it TEE-PLC}, two proof-of-concept systems that use ARM TrustZone TEE to protect the integrity of PLC's control logic execution and implement the prototypes focused on the PLC's control loop using open-source software, namely OP-TEE secure OS~\cite{op-tee} and OpenPLC~\cite{openplc}, on Raspberry Pi (Model 3B). {\color{black}The implementation is made available as open-source platform for future research and evaluation\footnote{Open-source project URL: \url{https://github.com/smartgridadsc/tee-plc/}}}.
    \item Evaluate the practicality of TEE-PLC and discuss desired features in ARM TrustZone TEE and secure OS implementation for future support of PLCs and other ICS devices.
\end{itemize}

We are aware that the SoC of Raspberry Pi 3B has a known security issue in the memory subsystem, and attackers can launch DMA (direct memory access) attacks to compromise ARM TrustZone~\cite{Stajnrod:2022}. However, our purpose is to evaluate the practicality and limitation of the control loop implemented by TEE-based PLCs under realistic use cases, not to deliver a full-fledged implementation that incorporates all security features provided by TEE or a product-ready system. Moreover, the discussed design can be reproduced on other embedded platforms. We also admit OpenPLC does not completely replicate the software running on commercial PLCs, but its process flow (e.g., scan cycle) and functionality are equivalent to commercial ones and are often utilized by researchers and engineers as a low-cost alternative~\cite{openplc2023}. Therefore, it still serves our purpose.    

The rest of the paper is organized as follows. We first provide the necessary background and a general model of a modernized PLC in Section~\ref{sec:bg}, where we also briefly introduce ARM TrustZone technology that we use throughout this paper. Next, we present the threat model and attack vectors against the integrity of PLC's control logic execution as well as define our security goals in Section~\ref{sec:model}. Then, we introduce two TEE-PLC designs of different levels of complexity and security guarantee 
in Section~\ref{sec:design}. In Section~\ref{sec:impl}, we elaborate on the TEE-PLC implementation using OpenPLC and OP-TEE secure OS. Section~\ref{sec:eval} presents performance evaluation and resource consumption metrics using the prototype. We then discuss the remaining challenges and recommendations for secure OS for ICS devices
in Section~\ref{sec:disc}. Related work is discussed in Section~\ref{sec:relatedwork}, and we finally conclude the paper in Section~\ref{sec:concl}.


\section{Background}
\label{sec:bg}

This section covers the overview of industrial control systems and PLCs to provide the context for the rest of this paper. We also discuss key features of ARM TrustZone that we are going to use for our proposed TEE-PLC design.

\subsection{Industrial Control Systems (ICS) Overview}

ICS is a general term that encompasses several types of control systems, including supervisory control and data acquisition (SCADA) systems, distributed control systems (DCS), and other smaller control system configurations, such as PLCs. ICS is deployed in a number of locations in our nations' critical infrastructure, including power grid systems, oil and gas plants, water treatment and distribution systems, transportation, manufacturing facilities, etc. 

To fulfill demands on stability and real-time responses in these settings, electronic and mechanical equipment in field sites are well-organized and inter-connected to coordinate with each other by using ICS communication protocols, such as Modbus, DNP3, OPC UA, Ethernet/IP, IEC 61850, and IEC 60870. Sensors are placed to perceive the operation status of industrial processes, and actuators take the next action based on control signals generated by PLC. 

\subsection{PLC Overview}
\begin{figure}
    \includegraphics[width=0.9\linewidth]{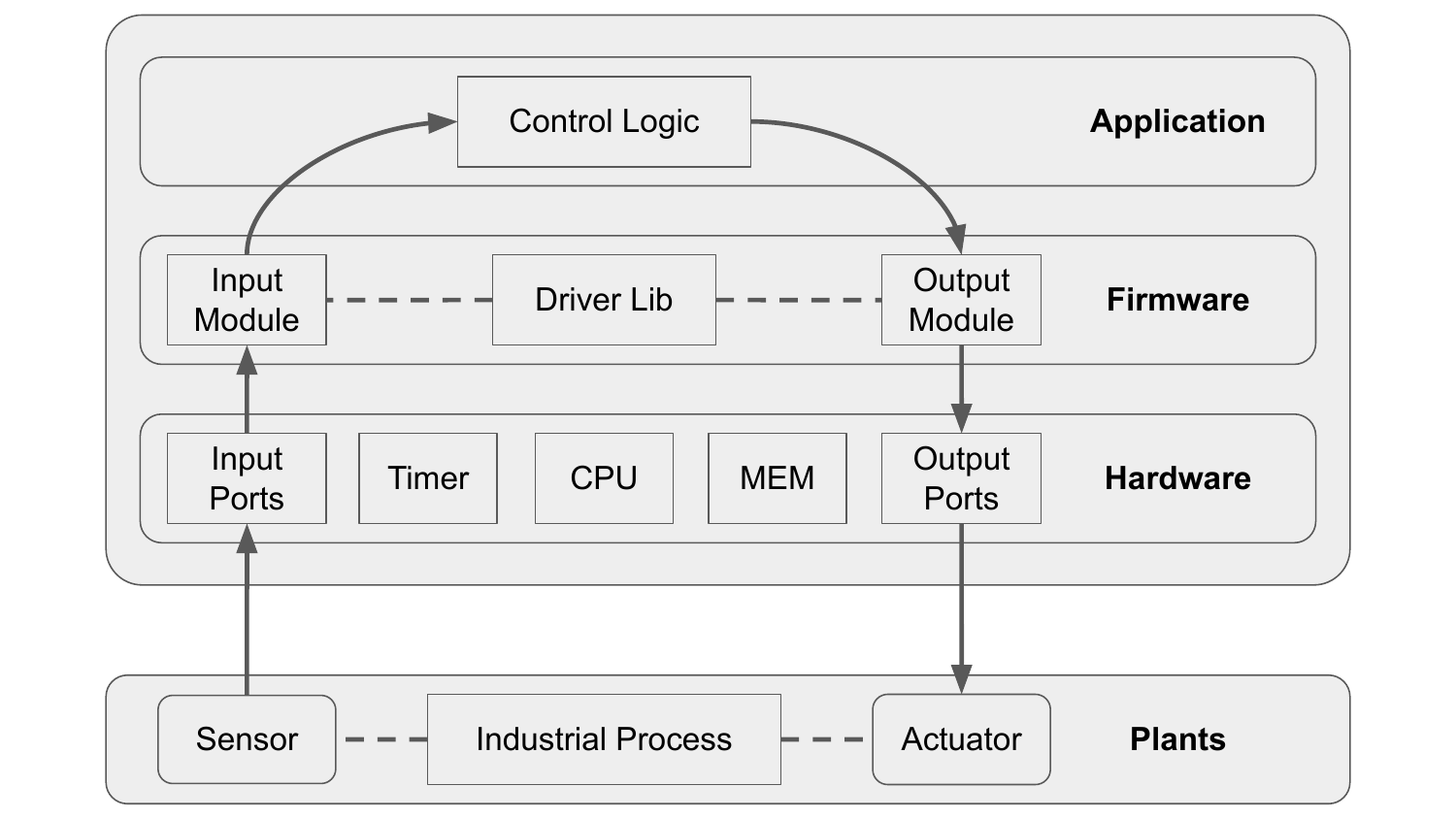}
    \vspace{-2mm}
    \caption{PLC Architecture \& Scan Cycle}
    \label{fig:plc_arch}
    \Description{PLC can be roughly divided into 3 parts: applications, firmware and hardware. Control logic runs in the application layer; I/O drivers run in the firmware layer; I/O ports work in the hardware layer. Each scan cycle starts by receiving input data from sensors, then carries out control logic execution and ends by sending output data to actuators.}
    \vspace{-4mm}
\end{figure}

PLC is a dedicated microcontroller (or microprocessor) with peripheral circuits fulfilling industrial-grade requirements. It generates output signals to energize actuators based on input signals collected from sensors. PLC can also collect runtime data and forward it to a control center. Based on that, the system operator can grasp what is going on in the physical plants. A typical PLC architecture can be divided into three layers from top down, shown in Figure~\ref{fig:plc_arch}. The application layer and firmware layer are the two software layers running upon the PLC hardware layer.

{\color{black}Control logic refers to the program for automated control running inside a PLC~\cite{ghaeini2019patt,yoo2019control,yoo2019overshadow}.} It consists of 4 different types of blocks: Instruction blocks, data block, configuration block and information block. Instruction blocks, as the name suggests, contain all the instructions running on PLC. The data block stores all the variables to be used in instruction blocks, like input, output, timer, counter and etc. Configuration block contains information about the entry address and size of each block, as well as the IP address for each sensor. The information block contains additional structural information about control logic and is used by engineering stations to recover readable code from de-compiled binary code. 
Process engineers develop industrial process control logic by using PLC programming languages standardized in IEC61131-3~\cite{iec61131}. 
IEC61131-3 specifies the syntax and semantics of programming languages for programmable controllers, including textual languages, such as Structured Text (ST).
%
To deploy control logic to production, process engineers upload the compiled control logic binary to PLCs. PLC runtime then loads the binary to memory and executes the control logic in an infinite loop while fulfilling the real-time requirement.

Firmware containing underlying OS and drivers of PLC peripherals for control logic to interact with the physical world. At PLC runtime, PLC firmware converts sensor signals received by PLC input peripherals and updates the input memory. Then a PLC executes the corresponding control logic based on input variables and updates the output memory. Finally, PLC firmware sends out control signals through output peripherals to actuators. This procedure, usually called a scan cycle, is executed repeatedly in fixed intervals.

\subsection{ARM TrustZone}

Modern operating systems and applications usually have huge sizes and still grow rapidly to support new features. However, the bloating code base also makes them vulnerable to various kinds of attacks. Troubleshooting and patching software with millions of lines of code can be time-consuming. Trusted execution environment (TEE), instead, provides a secure environment with minimal attack surface to manage critical information and data of applications. It maintains a minimal trusted computing base (TCB) and acts as a trust anchor for operating systems to complete different tasks requiring high-level security. 

\begin{figure}
    \centering
    \includegraphics[width=0.9\linewidth]{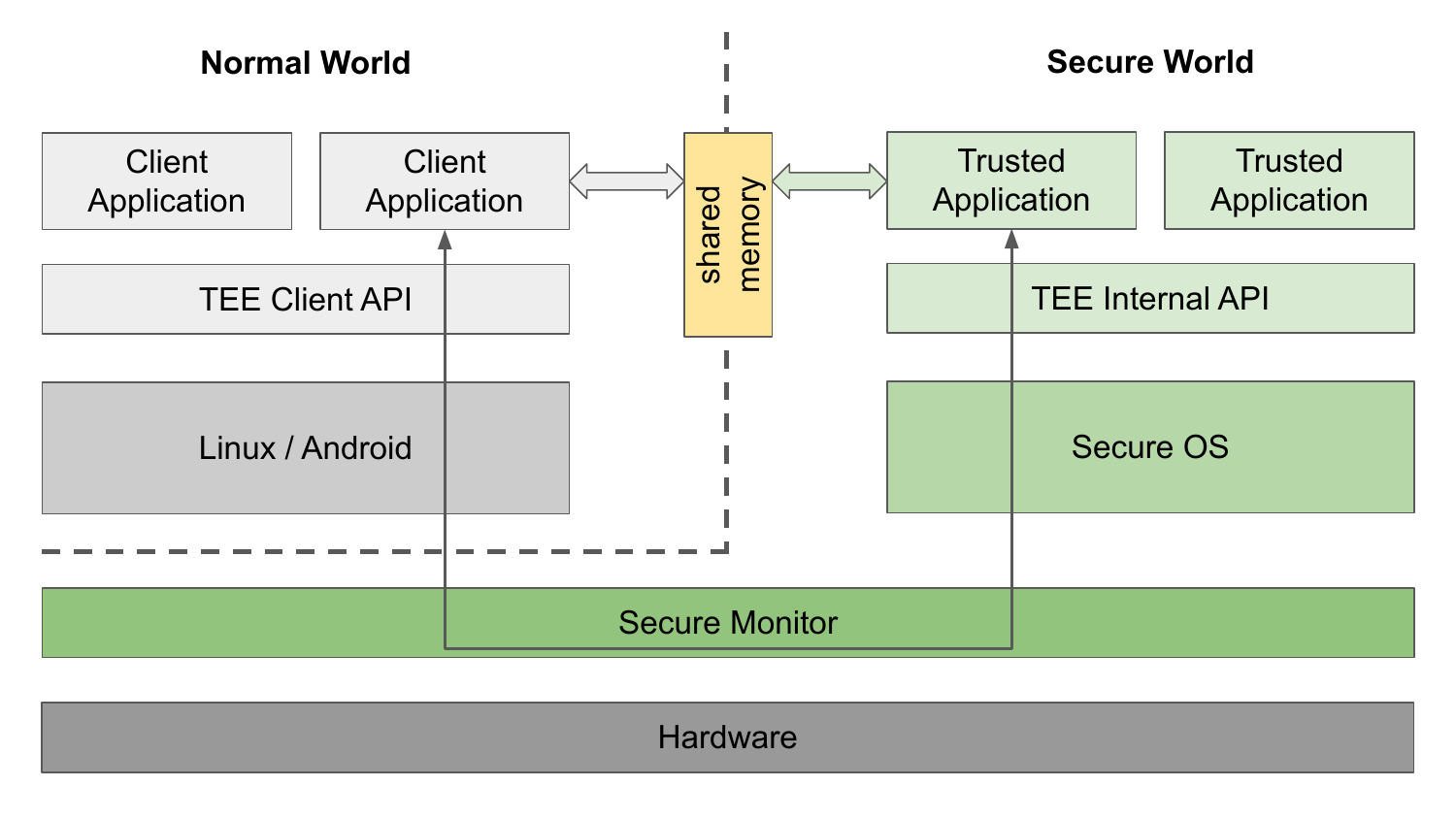}
\vspace{-3mm}
    \caption{ARM TrustZone TEE Architecture}
    \label{fig:trustzone}
\vspace{-3mm}

\end{figure}

ARM TrustZone technology~\cite{trustzone} is a hardware-enforced isolation mechanism to support the implementation of TEE. ARM introduced TrustZone-A for micro-processors since ARMv6-A and TrustZone-M for micro-controllers since ARMv8-M. The discussion in this paper is based on the former since it is more established and is employed by the products in the market already. TrustZone-enabled CPU divides its execution environment into normal world and secure world. Commodity OS and normal applications are running inside the normal world, whereas secure OS and trusted applications (TAs) are running inside the secure world. An overview structure of TrustZone is shown in Figure~\ref{fig:trustzone}. GlobalPlatform~\cite{gp2018internalapi} defines a series of TEE internal APIs similar to the Portable Operating System Interface (POSIX) standard to ease TA development. It also specifies the TEE Client APIs to ensure TA invocation by its corresponding Client Application (CA) in the normal world. However, more advanced features commonly used in Linux are not supported by secure OS, such as user-space threads and file systems.

%
ARM TrustZone technology has the following features to tackle security challenges for embedded devices such as PLC.

\noindent\textbf{World Switch and Shared Memory:} At any moment, either normal world or secure world is active. Switching between the two worlds is usually done by a handler called Secure Monitor running in the secure world. When the software in the normal world needs a specific service in a certain TA, it issues a Secure Monitor Call (SMC) to notify the CPU to suspend the workflow in the normal world and execute the nominated task in the secure world. When the task is done, SMC is called again to switch back to the normal world and resume the workflow in the normal world. ARM TrustZone guarantees that only necessary data is shared between the two worlds and other information is concealed or wiped out during world switch to avoid data leakage. To allow extra data exchange, TA developers can register shared memory between the normal world and the secure world. Abuse of shared memory can introduce vulnerabilities to the secure world; therefore, TA developers need to carefully define which data can be shared with the normal world through shared memory.

\noindent\textbf{Access Control of Memory and Peripherals:} The NS-bit of the CPU demonstrates the current execution environment and is propagated to debugging registers, caches, memory management unit (MMU), memory, interrupt controllers, and peripherals. All peripherals and memories can be configured to be secure or non-secure in the secure world. Non-secure tasks are not allowed to access secure peripherals or secure memories. When a malicious task running inside the normal world tries to read or write data inside secure memory, the CPU will raise an illegal access exception. This access control mechanism also prevents normal world tasks from configuring interrupt controllers or debugging registers to manipulate the control flow of the device.

\noindent\textbf{Secure Boot:} Boot process of embedded devices usually involves several boot images. Secure boot is a mechanism to protect all boot images from being tampered~\cite{arm2020secureboot}. It consists of both offline signing and online verification processes. Pairs of private key and public key are generated for signing and verifying images, respectively. During boot time, the image of the former boot stage will check the integrity and authenticity of the image for the next stage. A single signature verification failure terminates the whole booting process. This boot sequence establishes a chain-of-trust (CoT) from the very first boot image in ROM and guarantees the subsequent boot images and secure OS are authenticated and trusted. 

\noindent\textbf{Secure Storage:} Secure Storage allows TAs to read, store, and update critical data safely. Secure storage ensures integrity, confidentiality, authenticity, and atomicity. The content inside secure storage is only visible to the specific TA that creates it. This feature can be applied to the storage of private keys and critical configuration parameters of embedded devices.

 \section{Threat Models and Security Goals}
\label{sec:model}
\subsection{System Model}
In this paper, we focus on a field outstation in an ICS (e.g., a substation in a smart grid system), where PLCs are typically deployed, even though the proposed solution can be applicable to any PLC application.
In this section, we elaborate on the system model, using a modernized substation system in a smart grid system as an example, to discuss attack surfaces and threat models.

A substation is responsible for monitoring and control part of the power grid infrastructure~\cite{mashima2023cybersecurity}. Each substation consists of a number of sensors and actuators connected to the physical power grid components, such as circuit breakers, transformers, and transmission lines, for real-time monitoring and control. In the modernized substation, sensors and actuators are equipped with computation and communication capabilities, called intelligent electronic devices (IED) in the IEC 61850-compliant substation, and communicate with each other for coordination. While IEDs themselves implement some protection logic, such as over-current/voltage protection, higher-level control logic that involves multiple IEDs is often implemented on PLCs. PLCs periodically collect measurements from IEDs by using standardized ICS network protocol (e.g., Modbus and IEC 61850), and the resulting control command is also sent out to IEDs over the network. SCADA HMI may be deployed in the substation for situation awareness and manual control by human operators. PLCs are configured in an engineering workstation located in each substation or in the control center (connected over a wide-area network). Besides, outstations are often equipped with a VPN interface, which is utilized for remotely accessing the SCADA HMI, historian database, and/or directly interacting with ICS devices in the outstation (e.g., sending control/interrogation commands, updating firmware and logic on PLCs). 

\subsection{Attack Surface and Threat Models}\label{sec:threatmodel}
In this paper, we focus on attacks against the integrity of automated control logic execution by PLCs.
In other words, the ultimate attack goal is to manipulate the PLC to send incorrect control signals to actuators and cause the industrial process to fall into an unexpected state.
{\color{black} Besides, while real PLCs typically consist of input / output (I/O) modules for interacting with external devices, in addition to the processor and memory, we focus on the protection of data and logic run on the processor and memory, assuming the I/O modules are not compromised.}
%
%
We assume other ICS devices, such as sensors and actuators, are not compromised. as their security is orthogonal to the scope of this paper. 
In other words, we assume that the measurement data collected, locally stored, and sent out by sensors are accurate and that actuators follow the received control commands correctly. Those devices typically require a dedicated cable connection for configuration, and thus remote attackers in our scope cannot easily compromise, which supports the assumption.   

{\color{black}
While we are aware that a few attacks against TEE are recently reported (e.g., ~\cite{Stajnrod:2022}), in order to focus on the design and practicality evaluation of PLC scan cycle secured by ARM TrustZone TEE, we assume that ARM TrustZone TEE and the hardware needed by TEE are trusted. 
In other words, we assume that ARM TrustZone operates well and secure boot can load the secure kernel and secure services into secure memory and isolate from non-secure memory. 
Attacks against TEE are orthogonal to the contribution of this paper. To mitigate the risk, for instance, we can rely on a technology like vTZ~\cite{hua2017vtz}, which offers an additional layer of isolation for the secure world. On the other hand, the non-secure OS and all user-space applications in the normal world may be compromised by attacks.
}

We consider external attackers who are penetrating into the control system over the network, either through the VPN interface over the public network or the wide area network for SCADA communication among the control center and outstations, which can be either public or private network and also wired or wireless (e.g., cellular) network. For instance, attackers may exploit security flaws of the VPN server, such as a weak/compromised login credential (e.g., \cite{case2016analysis}), or known vulnerabilities like ShellShock~\cite{shellshock} for OpenVPN. Once it gets hacked, attackers gain access to the control system network. 
PLCs usually implement SSH and/or web-based user interfaces for configuration, and we assume such interfaces could be compromised by attackers, after which attackers could exploit the vulnerability for privilege escalation (e.g.,~\cite{plc-vuln}).
On the other hand, we exclude attackers who have physical access to the ICS  network or devices from our scope. We assume the facility is physically well protected to effectively counter physical penetration.
We also assume that 
engineering workstations for configuring PLCs and the insiders operating them are trusted.

\subsection{Security Goals}
\label{sec:goals}

\begin{figure}[!t]
  \includegraphics[width=0.9\linewidth]{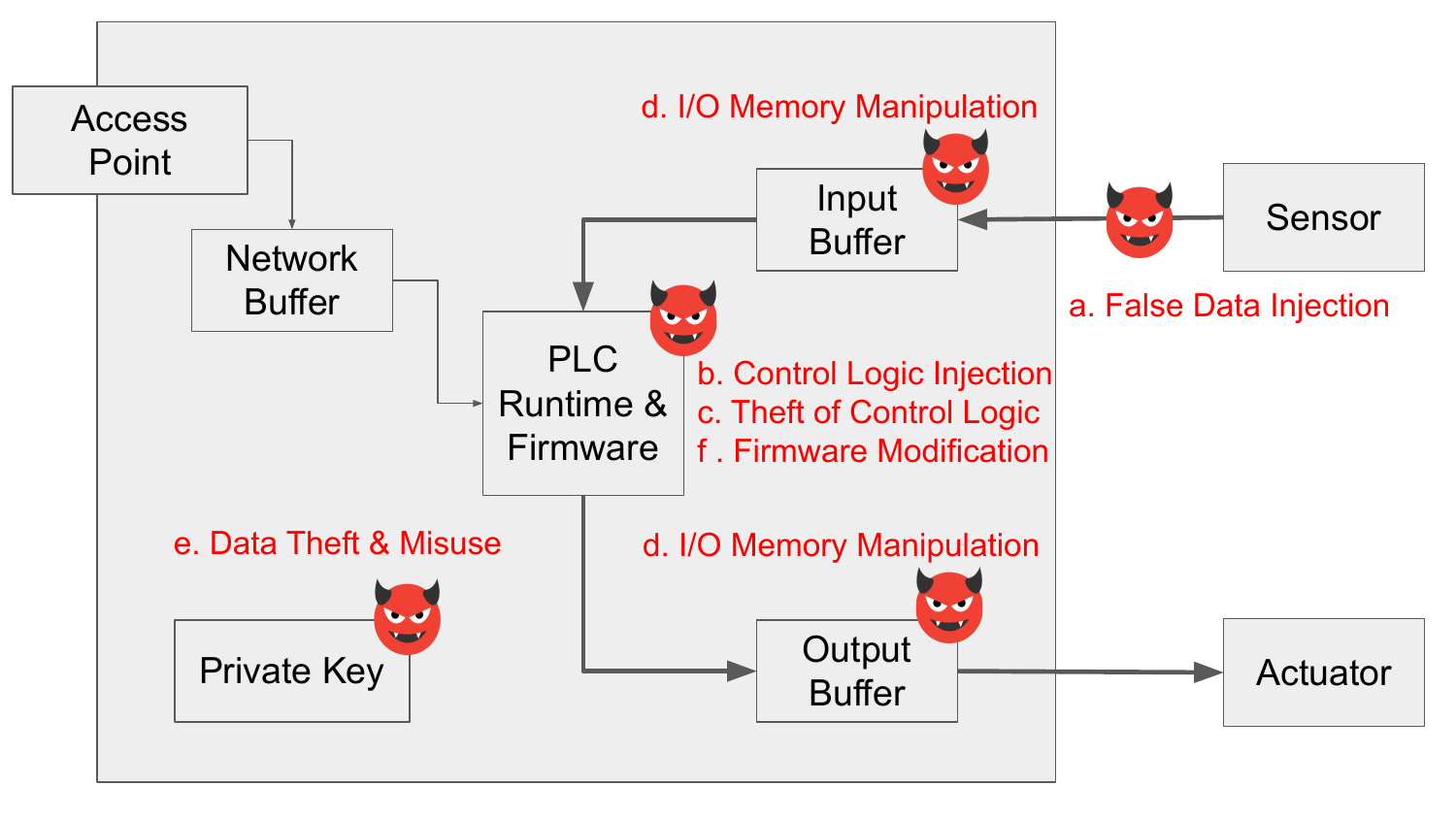}
  \vspace{-3mm}

  \caption{Data Flow \& Threat Model against PLC Scan Cycle}
  \Description{Data flow in field network}
  \label{fig:sys_model}
  \vspace{-4mm}
\end{figure}

Based on the discussed system and threat models, let us first discuss possible attack vectors. 
The high-level internal architecture and data flows in PLCs as well as attack vectors in our scope, are summarized in Figure~\ref{fig:sys_model}.
Our security goal is to mitigate the attack vectors we listed else and ensure the integrity of automated control by PLCs. 


\noindent \textbf{(a) False Data Injection:} Attackers who gained a foothold in the field outstation network can impersonate a sensor and send fake data to a PLC. This can be done, for instance, by means of Man-in-the-Middle attacks. Such attacks are possible because of the lack of message authentication. A successful false data injection can mislead the PLC to generate wrong control signals to actuators.

\noindent \textbf{(b) Control logic Injection:} Attackers with network access to a PLC can launch buffer overflow attack, etc., to inject malicious code in the control logic of the target PLC~\cite{yoo2019control}.
This may also be achieved by exploiting flaws in access control, such as weak/default credentials to gain initial access to the PLC and defect in buffer unbounded checks to send a malicious buffer to the PLC. The injected code would cause the PLC to execute malicious code snippets and thus can sabotage the physical plant or suppress control actions by changing threshold values in the logic.

\noindent \textbf{(c) Theft of Control Logic:} An attacker who has access to the PLC device via the network may attempt to steal the binary or plain-text code of the control logic. Details of the control logic could provide hints for attackers to understand the system configuration. Moreover, PLC codes are also utilized to design intrusion detection rules~\cite{tan2022cotoru}, and thus knowledge about the PLC logic could potentially reveal the configuration of cybersecurity measures deployed.   

\noindent \textbf{(d) I/O Memory Manipulation:} 
Attackers may gain root access to a PLC by exploiting vulnerabilities of OS or firmware. Such an attacker can further launch a malicious program to persistently change the I/O memory of the PLC and interfere with proper data receiving and sending~\cite{abbasi2016ghost}. 
{\color{black}
An attacker could also mount a data-oriented attack, e.g., by means of buffer overflow, to affect the critical variables that affect the control flow of the PLC logic. Such an attack is possible even by attackers on the network.
}
Attacks of this sort would result in false data injection and sending of inappropriate control commands to actuators.

\noindent \textbf{(e) Data Theft and Misuse:} After gaining root access, an attacker could compromise private keys and credentials in the PLC's file system, which allows the attacker to impersonate the legitimate PLC and send crafted packets to other devices to mislead them to generate incorrect control signals.

\noindent \textbf{(f) Firmware Modification:} Attackers with root privilege can force a control logic update or firmware replacement. Malicious firmware may contain rootkit and launch stealthy attacks like HARVEY~\cite{harvey}.

To protect PLCs from being compromised by the attack vectors mentioned above, existing solutions, for instance, ones based on cryptographic primitives, are not sufficient.
Although secure communication schemes such as TLS/SSL can establish a trusted channel and counter false data injection attacks on the network, there is still a possibility of manipulation of data in the PLC. 
Once an attacker has root access to the PLC, any sensitive data stored on the storage is no longer secure, even if they are encrypted.
Ultimately, once the firmware is compromised by the attacker, no security scheme running on top of it can be fully effective. 

Even if attackers gain access to PLC and obtained certain privileges, we need control logic integrity and execution of it to be still trustworthy. Trusted Execution Environment (TEE) technology can fill these gaps by sealing private key insecurity storage and protecting the collection of sensor data and control logic execution in the isolated execution environment. 
Given that many PLC products are ARM-based, we focus on ARM TrustZone technology among the TEE technologies currently available in the market because it is more likely to be adopted in the near future.




\section{TEE-PLC Design}
\label{sec:design}


We next discuss our approach to counter the aforementioned attack vectors using ARM TrustZone TEE.  
{\color{black} 
Since our focus in this paper is the evaluation on the feasibility and practicality of the PLC scan cycle secured by TEE through a proof-of-concept implementation, we don't aim at full-fledged PLC implementation. Moreover, some of the TEE's security features, such as secure boot, are assumed but not specifically emphasized in our proof-of-concept designs since it is not essential for our evaluation.}

{\color{black}
\subsection{Preliminary Design}

When designing a PLC on TEE, an important decision is which modules/components of the PLC to deploy in the secure world. The components in the secure world are regarded as the trusted computing base (TCB), and thus it should be minimal. On the other hand, when we consider practical usage, the number of world switches should be minimal to avoid high overhead for the sake of real-time operation. It is also crucial that the update of the PLC logic can be done easily without requiring modification on other parts of the PLC runtime. 
Our main objective in this paper is to secure the integrity of the PLC scan cycle, which consists of the collection of sensor data, execution of logic, and triggering control when a certain condition is met.  
A naive option we could immediately consider is to deploy the control logic execution in the secure world as a TA. We could 
sub-divide the PLC control logic into multiple modules and deploying them in normal and secure world depending on the criticality to shrink the TCB further. For instance, we could conduct static analysis to identify critical variables and could deploy segments of logic that touch them in the secure world. However, this approach is expected to add complexity (e.g., an increase of entry points for the secure-world modules) and result in an increase in the number of world switches, which is very costly. Thus, we do not explore such a direction and treat the whole control logic as the minimal unit to be protected in the secure world.

\begin{figure}[!t]
\vspace{-2mm}
    \centering
    \includegraphics[width=0.7\linewidth]{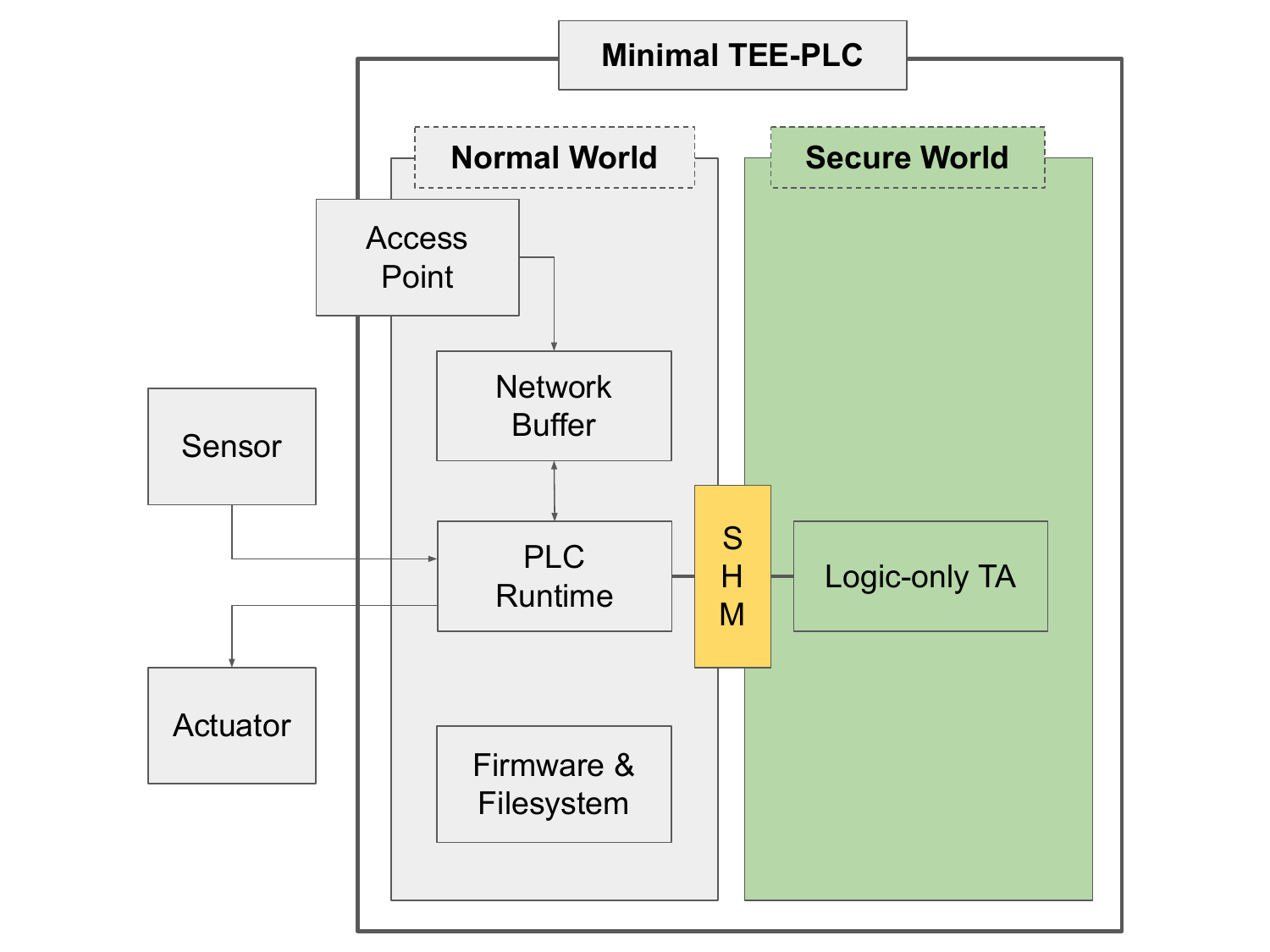}
    \vspace{-2mm}
    \caption{Minimal TEE-PLC Architecture}
    \Description{Control logic is extracted from the normal world and compiled into TA running inside the secure world; Secure Communication Module is implemented as Modbus/TLS client.}
    \label{fig:minimal_tee_plc_arch}
    \vspace{-4mm}
\end{figure}

In this design, which we call {\it Minimal TEE-PLC} (Figure~\ref{fig:minimal_tee_plc_arch}), the PLC runtime in the normal world takes care of the communication with sensors for data collection and at a certain interval, invokes a TA that implements PLC control logic (called {\it Logic-only TA}), via a secure monitor call (SMC) and passes the collected data via the shared memory. 
The rest of the PLC runtime (e.g., communication with other devices) and user interfaces are left in the normal world. 
Then the PLC control logic is executed in the secure world, and the outcome is passed back to the PLC runtime in the normal world to send out control commands if necessary.
This way, we can protect the integrity and confidentiality of the PLC control logic since it is stored and executed in the secure world. 

Because PLC logic often requires an update after the deployment of the device, it is crucial to have a secure interface to allow the installation of only TAs (or updates of them) that are authorized by the infrastructure operator.
Although discussion on specific technologies is outside of our scope, we assume TA codes are verified, before authorization, using methodologies like static/dynamic analysis by the TA developer to ensure that there is no security vulnerability.
Otherwise, a malicious module would be installed, which mounts memory and/or control flow attack against the Logic-only TA in the secure world. 
Although this is outside of our prototype, for such secure TA update in ICS, we can rely on GateKeeper framework~\cite{gatekeeper}.
In short, GateKeeper utilizes a ``package manager'' TA in the secure world that not only enables installation or update of a TA flexibly without secure OS vendor support but also is responsible for enforcing cryptographic verification (integrity and authenticity via a digital signature) before installation.
Since we have implemented the PLC control logic as a single TA as discussed in this section, control logic binary distribution can follow the secure TA distribution process proposed in Gatekeeper~\cite{gatekeeper}.
In this TEE-PLC design (and also the enhanced version discussed later), the update of control logic on PLC is replaced with an update of the Logic-only TA, which contains all codes and data needed for control logic execution. 

However, false data injection and I/O memory manipulation cannot be countered. For instance, if the data is manipulated or fake data is injected into the network, owing to the lack of secure communication, the attack cannot be detected or prevented. Recently the use of TLS for ICS communication is recommended by international standards like IEC 62351~\cite{iec62351}. Unfortunately, even if secure and authenticated communication is implemented, it is not fully effective as long as it is implemented in the normal world. Namely, the is a window of vulnerability between the decryption/verification of messages and processing at the Logic-only TA. For instance, when the data is being passed to the secure world using the shared memory, an attacker who has control of the device (i.e., normal world) can manipulate it. The assessment of this design against the attack vectors is summarized in Table~\ref{tab:security_mapping}.

\begin{table}[h]
\centering
\color{black}
\caption{Security Goals and TEE-PLC Design}
\label{tab:security_mapping}
\vspace{-2mm}

\begin{tabular}{|l|c|c|}
\hline
Attack Vector & \pbox{1.5cm}{Minimal \\  TEE-PLC} &
\pbox{1.5cm}{Enhanced TEE-PLC}\\
\hline
\hline
(a) False Data Injection & & \checkmark \\ \hline 
(b) Control logic Injection & \checkmark&\checkmark \\ \hline
(c) Theft of Control Logic & \checkmark& \checkmark\\ \hline
(d) I/O Memory Manipulation & & \checkmark\\ \hline
(e) Data Theft and Misuse & (N/A) & \checkmark\\ \hline
(f) Firmware Modification & Secure Boot & Secure Boot\\ 
\hline

\end{tabular}
\vspace{-4mm}
\end{table}

}

\subsection{Enhanced TEE-PLC Design}
\begin{figure}[!t]
\vspace{-1mm}
    \centering
    \includegraphics[width=0.9\linewidth]{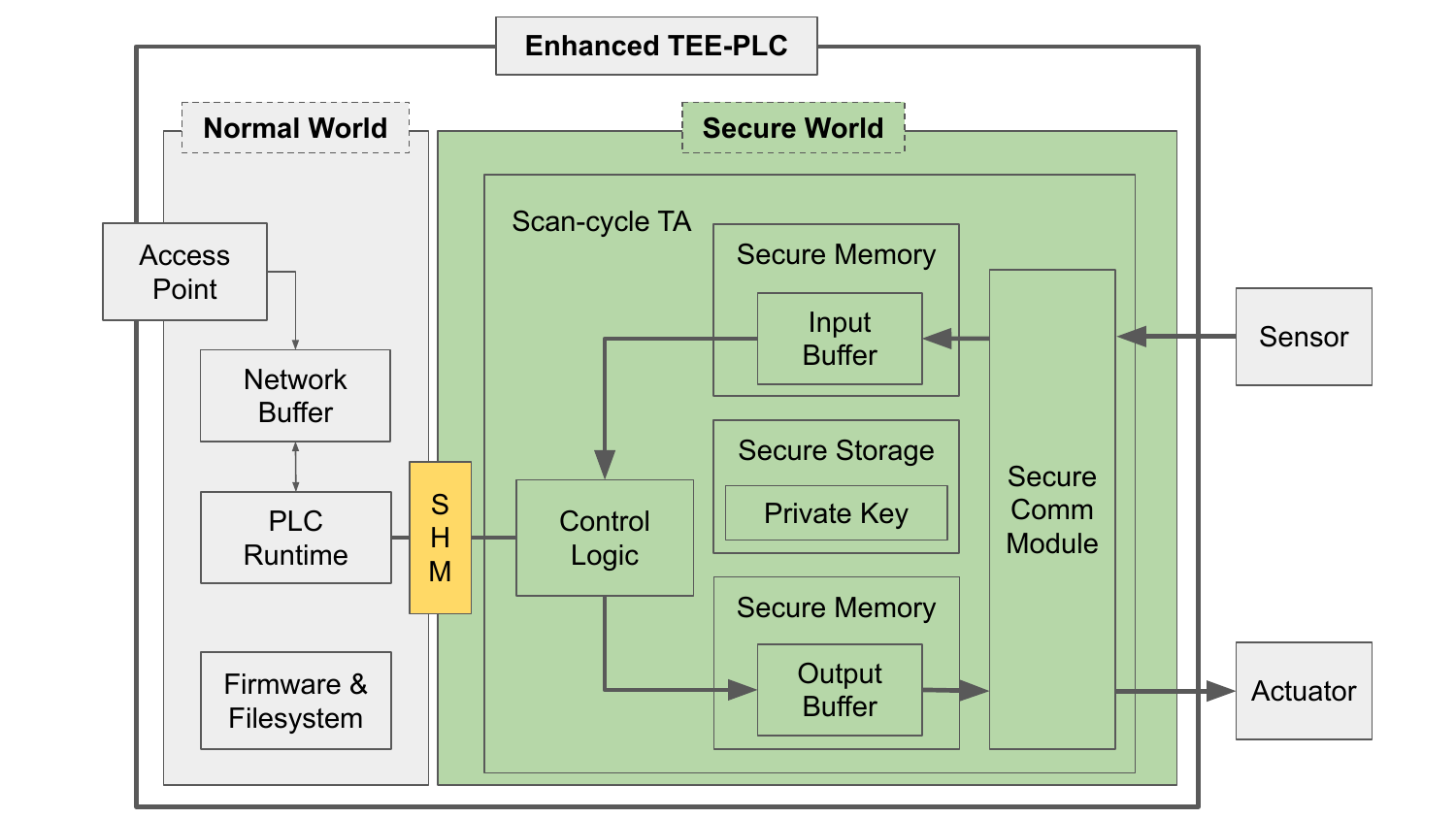}
    \vspace{-2mm}
    \caption{Enhanced TEE-PLC Architecture}
    \Description{Control logic is extracted from the normal world and compiled into TA running inside the secure world; Secure Communication Module is implemented as Modbus/TLS client.}
    \label{fig:enhanced_tee_plc_arch}
    \vspace{-3mm}
\end{figure}

%
%
{\color{black}
To address the limitation of Minimal TEE-PLC design, we can consider deploying PLC communication with sensors and actuators also in the secure world as part of the TA. In this design, we can counter I/O memory attacks mounted in the normal world because the collection of sensor measurements and logic executions are both done atomically in the secure world.  
%
According to the GlobalPlatform specification~\cite{gp2021teesocket}, TEE\_iSocket offers a TCP/IP network interface in the secure world. However, the actual network stack relies on the implementation in the normal world. 
This implies that if the normal world is compromised, messages could be under attackers' control. Thus, we need to do message encryption and authentication in the secure world. 

Fortunately, standardized communication protocols in the industrial control system offer options to protect communication by using TLS. For instance, in the smart grid domain, IEC 62351 standard recommends using TLS for securing any communication over TCP/IP, including IEC61850, DNP3, and Modbus~\cite{hussain2019review,iec62351}. Thus, we implement the TLS communication module as part of the TA in our prototype. 
By establishing an end-to-end secure communication channel (such as TLS with mutual authentication) between the TA in the secure world and external devices, the false data injection and manipulation of outgoing messages can be countered.   
%
%
%
However, this design decision implicitly assumes that sensors and actuators also support TLS, which is unfortunately not always the case in real-world ICS deployment. In such a situation, we can employ a bump-in-the-wire approach (e.g.,~\cite{esiner2019f,castellanos2017legacy,tippenhauer2021vbump}) for secure communication, and the algorithm implemented in the secure communication module in the TA can be replaced accordingly.
Secret information needed for secure communication, such as a private key and a shared secret key, can be stored in the secure world and thus protected from the attackers outside of the secure world.  

We call the enhanced TA {\it Scan-cycle TA}, and in summary, its functionality includes the following:
\begin{enumerate}
\item Poll data from external sensors via secure communication 
\item After collecting all necessary data, execute the control logic
\item Trigger control commands to actuators using a secure communication channel
\item Make the updated measurements and status available to the normal world via shared memory (for the user interface as well as queries from SCADA HMI)
\end{enumerate}


When implementing the TA, we need to consider the single-threaded nature of TEE as well as the cost of the world switch. The former means that the TA cannot be multi-threaded and also that during the execution of the secure world module, the normal world is paused. TEE-PLC implementation should consider these factors to meet the real-time demand of the PLC functionality.  
In this paper, we choose the most straightforward way of implementation to minimize the world switch and functionality to be deployed in the secure world. Namely, all of the 4 tasks above are executed sequentially once the Scan-cycle TA is invoked. 
While we don't claim this is the optimal design, we think it serves as the baseline for further enhancement and optimization study. 

In the design using the Scan-cycle TA, the PLC run-time in the normal world invokes the TA periodically using SMC. At the invocation, the PLC runtime does not have to pass any data to the TA since all the tasks for the scan cycle are implemented in the TA, and necessary configurations are stored in the TA. After the execution of the scan cycle, the sensor data collected by the TA are passed to the PLC runtime in the normal world via the shared memory so that the user interface of the PLC can access it.  

The design is summarized in Figure~\ref{fig:enhanced_tee_plc_arch}, and we call this implementation using the Scan-cycle TA {Enhanced TEE-PLC}. The flowchart can also be found in Figure~\ref{fig:data_flow} in Appendix A. Next, we discuss the security of the Enhanced TEE-PLC. 
}

\subsection{Security Discussion on Enhanced TEE-PLC}

In the Enhanced TEE-PLC design, we made a clear-cut separation of PLC functionality into two worlds. Only components needed by integrity-protected scan-cycle execution are placed in the secure world as a TA, while the rest, such as user and server interfaces, are left in the normal world. Besides, each scan-cycle execution is done via a single invocation of the Scan-cycle TA, which minimizes interaction and data exchange between the worlds. If there is any sensitive data (e.g., private or secret key) required for the logic execution, it can be stored in secure storage.
%
Next, we discuss how our design can accomplish our security goals against attack vectors discussed in Section~\ref{sec:goals}.


Authenticated and encrypted channels established by the secure communication module inside the secure world guarantee the integrity, confidentiality and authenticity of data flow between the PLC (more specifically, the Scan-cycle TA) and sensors/actuators. This effectively counters attacks not only on the ICS network but also in the normal world on the PLC. The sensor measurements are carried over from the secure and authenticated channel to the TA in the secure world, and thus injection and manipulation of messages by unauthorized entities in the network can be detected before logic execution. The outgoing control commands can also be protected.
Even when an attacker is on the device (the normal world), it cannot manipulate I/O buffers to pass data to the TA, since it is done in the secure world. 
Therefore, we can effectively counter the attack vector {\bf (a)} and {\bf (d)}.  
Although we have excluded the compromise of legitimate sensors from our scope, 
it is important to consider a case where 
a legitimate sensor is compromised and then sending fake data over the secure channel. 
Detection of such attacks would require physics-based attack detection (e.g., \cite{sourav2023machine}) run in the secure world. This is technically possible but results in bloated TCB. 


Control logic execution inside the secure world ensures both static and runtime integrity and confidentiality. To inject malicious code snippets to the control logic, for example, Stuxnet~\cite{stuxnet} uploads malicious code blocks to PLC memories through infected engineering software. Data execution and fragmentation \& noise padding attacks~\cite{yoo2019control} stealthily transfer crafted code segments by masquerading as legitimate data packets. Buffer overflow attacks inject code snippets into stack or heap buffer. 
In our design, control logic is implemented as a TA stored in the secure world, and by using GateKeeper~\cite{gatekeeper}, which prevents the installation of unauthorized TAs by means of a white list of authorized TAs and digital signatures, injection of unauthorized modules into the secure world is highly infeasible. 
%
A potential interface for an attacker to compromise the control flow integrity of the Scan-cycle TA is only the secure monitor call to invoke the TA. However, the arguments to be passed here are the universally unique identifier (UUID) allocated to the TA as well as the entry point identifier. These can be strictly type-checked and thus can be validated to avoid attacks such as buffer overflow. We should also note that the shared memory to pass data from the Scan-cycle TA to the normal world is used in a one-way manner. In other words, the TA is not reading data from the shared memory. Thus, even if this shared memory contains malicious data injected by the attacker, it does not affect the execution of control logic. 
Regarding attacks from the network, which aim at buffer overflow etc., they can be countered by the appropriate size and boundary checks implemented in the TA. 
Thus, the goal {\bf (b)} is addressed.

PLC logic is encapsulated in the Scan-cycle TA deployed in the secure world. Thus, an attacker in the normal world of the PLC device cannot access the logic. However, in practical application scenarios, PLC logic often needs to update after deployment, and such an update is performed over the network. Thus, in order to ensure the confidentiality of PLC logic, we should protect the confidentiality (and also integrity) of the PLC logic in transit. TEE-PLC scheme allows PLC logic update by updating the Scan-cycle TA. GlobalPlatform specification~\cite{gp2018internalapi} defines a way to install and update the TA securely, and according to the specification, the TA to be installed is integrity protected by digital signature and optionally encrypted. 
We rely on GateKeeper framework~\cite{gatekeeper} based on the TA installation/update mechanism to enable a flexible, operator-centric TA update framework to fit real-world use cases. 
It might be argued that the control logic could be inferred by monitoring the input and output messages. However, by encrypting the input and output messages, such an attack is made significantly difficult. 
Combining these, our design can counter the attack vector {\bf (c)}.  

In our design, any sensitive information that is necessary for the control logic execution, including private keys of PLCs, is stored in the secure storage in the secure world. Thus, the data at rest is not accessible to the attacker on the network or in the normal world of the device. Moreover, such data does not have to leave the secure world during the runtime of PLC. Thus, an attacker could not steal and abuse it for impersonation, addressing the attack vector {\bf (e)}.


Firmware rootkits like HARVEY~\cite{harvey} depend on the unprotected firmware update process. While it is not the contribution from our TEE-PLC designs, secure boot, which is one of the key security features provided by ARM TrustZone TEE, prevents attackers from modifying firmware by means of either remote firmware update or offline update through SD card. Any modification to the secure OS will cause boot failure. Furthermore, forcefully flashing new firmware through a JTAG (Joint Test Action Group) interface is also infeasible if the JTAG interface is correctly configured as a secure peripheral and JTAG debugging from the normal world is disabled. 

{\color{black}
In this section, we provided the qualitative security discussion under the threat models and assumptions discussed in Section~\ref{sec:threatmodel} (i.e., an ideal case where ARM TrustZone and secure world components are vulnerability free), to justify our design. A proof-of-concept implementation using open-source software, which aims at enabling practicality evaluation of such a TEE-based PLC, will be detailed next. 
Empirical security evaluation using the prototype would be possible. However, exploitation of vulnerabilities in the specific open-source framework we use would not provide much real-world insights and thus should be considered orthogonal to our contribution in this paper. 
}

\section{Prototype Implementation}
\label{sec:impl}
{\color{black}In this section, we elaborate the prototype implementation of Enhanced TEE-PLC. Owing to the limitation of space, we don't elaborate on Minimal TEE-PLC, but it is very similar, except that it does not have a secure communication module in the secure world.
}

We implemented our prototype of TEE-PLC on Raspberry Pi (Model 3B). This device has a 4-core Cortex-A53 CPU (ARMv8-A architecture) and 1GB RAM, supporting the latest Arm TrustZone technology. The commodity OS we use in the normal world is a TrustZone-aware Linux kernel (version 5.17). To better mimic the performance of real-world PLCs, we limited the Linux kernel to use only 256MB RAM as non-secure memory and bring up only 2 cores running at the fixed 600 MHz. We chose OP-TEE~\cite{op-tee} as the secure OS to run the Logic-only TA or Scan-cycle TA in the secure world. 32MB RAM is allocated as secure memory for OP-TEE and TA. OpenPLC~\cite{openplc} is chosen as the PLC runtime in the normal world 
invoking the TA in the secure world. 

OpenPLC~\cite{openplc} is an open-source PLC software that well approximates real PLC products and thus is widely used by industry experts and researchers for testbedding and experiments.
%
It integrates a complete toolchain for PLC development and deployment, including OpenPLC Editor, an IEC 61131-3 compliant logic programming environment, PLC runtime, and a web-based user interface. The web interface allows operators to configure PLC parameters, upload PLC code (in IEC 61131-3 Structured Text), and then start and stop the PLC runtime. It also allows users to browse the measurements collected from sensors. 
OpenPLC utilizes internally MatIEC~\cite{matiec} to compile IEC 61131-3 compliant PLC logic code into C code, which is compiled for execution.
%
It also supports commonly used industrial network protocols, such as Modbus and DNP3, to build I/O communication channels with sensors and actuators. 

OP-TEE~\cite{op-tee} is one of the most popular open-source implementations of the secure OS for ARM TrustZone. It has support for GlobalPlatform TEE Client API Specification v1.0~\cite{gp2010clientapi} and TEE Internal Core API Specification v1.1.2~\cite{gp2018internalapi}. The security features we discussed in Section~\ref{sec:bg} are also supported by OP-TEE. 
At the time of our implementation, the latest version of OP-TEE is 3.17.
We built the TA using internal core APIs. We also set up a TCP/IP client in the secure world using TEE\_iSocket APIs provided by OP-TEE. We further implemented a Modbus/TLS client as the secure communication module in the secure world using WolfSSL~\cite{wolfssl}.

The rest of this section discusses the technical details regarding the Enhanced TEE-PLC prototyping. In particular, we elaborate on the overall architecture, implementation of Scan-cycle TA, including secure communication, and required changes in OpenPLC, such as the interface to interact with Scan-cycle TA.
Our discussion on implementation focuses on OpenPLC, but the overall design (e.g., separation of the PLC runtime into normal and secure world modules and interface among them) is applicable to other PLC runtime implementation written in C that supports OP-TEE API.

\subsection{Scan-cycle TA}
\begin{figure}
    \centering
    \includegraphics[width=0.95\linewidth]{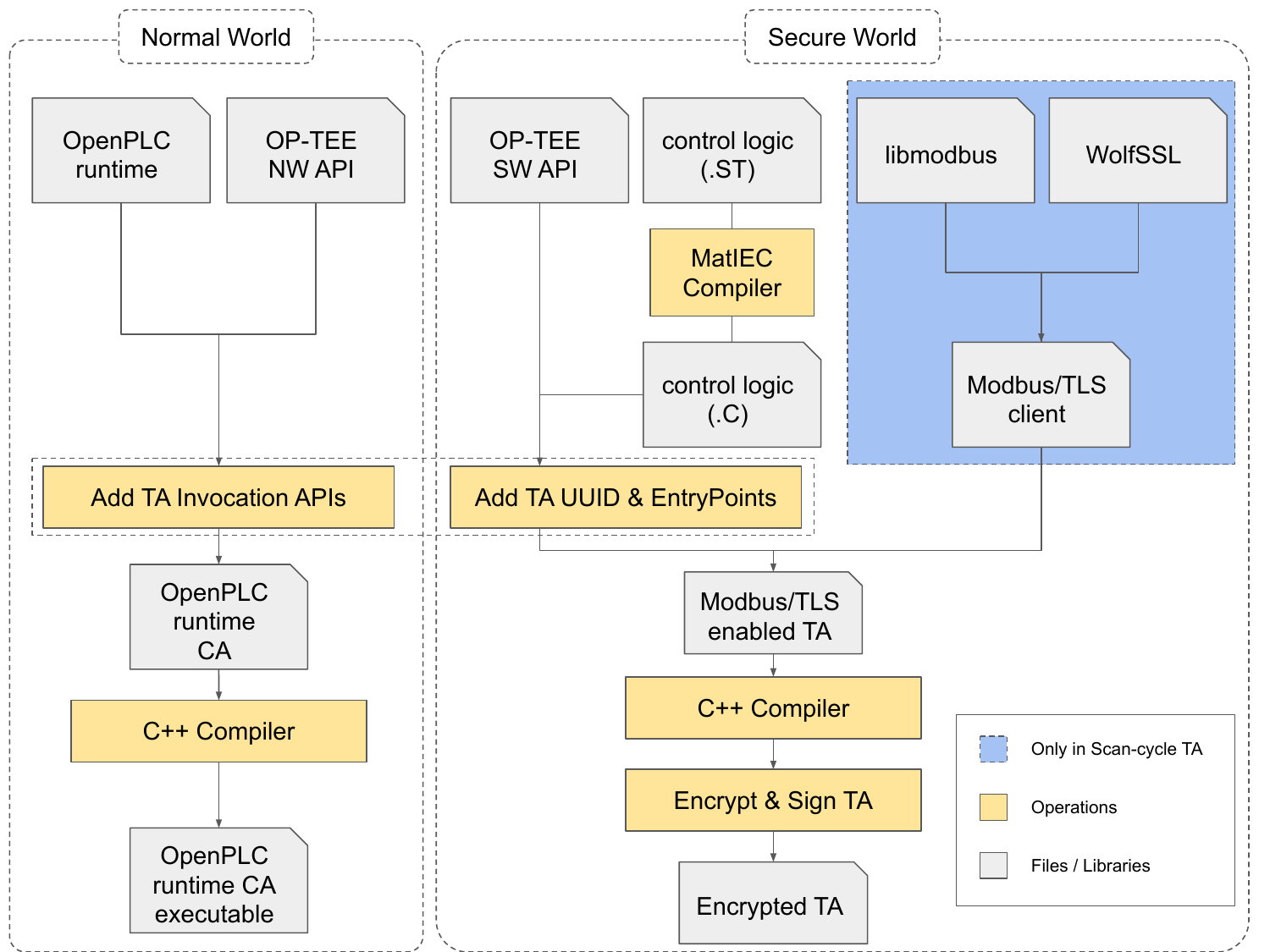}
    \caption{TEE-PLC Compilation Workflow}
    \Description{Control logic is extracted from the normal world and compiled into TA running inside the secure world; Secure Communication Module is implemented as Modbus/TLS client.}
    \label{fig:openplc_optee_workflow}
    \vspace{-4mm}
\end{figure}

%
Our Enhanced TEE-PLC design protects control logic execution by making it run inside the secure world. That means control logic is implemented as a standalone TA and needs to be invoked by OpenPLC runtime in the normal world when a scan cycle is triggered. Data sharing between the TA and the OpenPLC runtime in the normal world is also needed. The interfaces for mutual communication between the secure world and the normal world are specified by GlobalPlatform~\cite{gp2010clientapi,gp2018internalapi}. 
The right side of Figure~\ref{fig:openplc_optee_workflow} depicts the steps of how we create the Scan-cycle TA. First, we generate a UUID for the TA and define a series of entry points to expose secure services to the normal world (See also Figure~\ref{fig:data_flow} in Appendix A). 
Second, we implement callback functions required by OP-TEE. 
Third, we specify the size of the heap and stack of TA to ensure a minimum memory footprint; We also tag our TA with a version number to prevent downgrade attacks. When the PLC logic needs to be updated, a new TA with a higher version number and the same UUID can be developed and deployed on the TEE-PLC. TA file with a lower version will be rejected by OP-TEE~\cite{gatekeeper}. 


At this point, secure services in the TA can be invoked by OpenPLC runtime by specifying the UUID and the entry point ID. The next step is to add control logic to this TA template. We employ MatIEC to compile PLC control logic (written in Structured Text or ST file) into C code (\textit{POUS.c}, \textit{ResX.c}, \textit{ConfigX.c}). \textit{POUS.c} is the file containing all the code blocks (subroutines) of the PLC control logic, \textit{ResX.c} deals with the initialization and update of variables for each code block. \textit{ConfigX.c} arranges the control flow of all subroutines and provides an entry point to all the code blocks. MatIEC also generates \textit{LOCATED\_VARIABLES.h}, which contains declaration of I/O variables needed by control logic. We need to define these variables in Modbus client such that control logic can read or write slave data collected from or to be sent to slave devices.

We implement Modbus/TLS client as the secure communication module in the secure world. Modbus follows a query-response pattern, meaning each read or write operation involves a query packet from the Modbus client and a response packet from the Modbus server. In each scan cycle, the Modbus client in the secure world sends reading requests to sensors and writing requests (i.e., a control command) to actuators. 
We implement the Modbus client by porting libmodbus~\cite{libmodbus} to the secure world. The network communication is handled by the underlying TEE\_iSocket interface. To support secure transmission, we add a TLS layer using WolfSSL~\cite{wolfssl}. 
TLS handshaking and session management are also handled by WolfSSL. We port WolfSSL to OP-TEE so that it can work in the secure world. Meanwhile, the private key and certificates for TLS communication are stored in the secure storage. 

To query each slave, the TEE-PLC needs to know the IP address, port number, and slave number of the slave. These configuration can be stored in a separate configuration file or database. However, this can introduce extra complexity and overhead to the secure world. Therefore, we hardcoded the list of IP addresses and ports of such devices in the Scan-cycle TA since
the sensors and actuators to interact with are tightly coupled with the logic.

\subsection{Integration with OpenPLC Runtime}

To invoke a secure service of the Scan-cycle TA in the secure world, OpenPLC runtime in the normal world also needs minor changes, as shown in the left side of Figure~\ref{fig:openplc_optee_workflow}. The original control logic in the normal world is replaced with a series of OP-TEE client APIs to invoke the Scan-cycle TA, using the TA's UUID and secure services' entry point IDs. OpenPLC runtime also allocates a shared memory for collecting the latest measurement data from the Scan-cycle TA. Such data is used for displaying the information on the dashboard and also utilized when the PLC is queried by another device, such as SCADA HMI. 
The Enhanced TEE-PLC's
process flow 
is shown in Figure~\ref{fig:data_flow} in Appendix A.






\section{Evaluation}
\label{sec:eval}

While ARM TrustZone technology is promising to provide additional layers of defense for ICS devices like PLCs, one drawback we should take into account is overhead. 
On the other hand, ICS devices often have stringent latency requirements for real-time operation. Thus, in this section, we use the Minimal and Enhanced TEE-PLC prototypes discussed in the previous sections to evaluate the practicality. 
As found in Table~\ref{tab:security_mapping}, the Minimal version offers a weaker security guarantee. On the other hand, it has a smaller TCB and is expected to offer better performance because communication with sensors can be done in the normal world, and thus it incurs less overhead and can be well parallelized (i.e., less subject to the single-thread constraint in the secure world~\cite{wan2020rustee}).

\subsection{Experiment Setup}
We created a testbed simulating a real-world field network by connecting Raspberry Pi, sensors, and actuators to the same router through Ethernet cables. Sensors and actuators are implemented with Python using pyModbus~\cite{pymodbus} and Python SSL. 
In a single scan cycle, the Modbus client (i.e., the PLC) first sends $read\_coils$ to query input data from sensors and then sends $write\_coils$ to write output signals to actuators. According to $ping$ statistics, the round trip delay is around 1.2ms.

%

\begin{table*}[!h]
  \captionof{table}{Execution Time ($ms$)}
  \vspace{-2mm}
  \label{table:exec_time}
  \begin{tabular}{||c|c|c|c|c|c|c|c|c|c|c|c|c||}
     \toprule
     \multirow{3}{4em}{} & \multicolumn{12}{c||}{\textbf{Pairs of connected sensor \& actuator}} \\
     \hline
     & \multicolumn{3}{c|}{\textbf{1} (2 slaves)} & \multicolumn{3}{c|}{\textbf{2} (4 slaves)} & \multicolumn{3}{c|}{\textbf{4} (8 slaves)} &\multicolumn{3}{c||}{\textbf{8} (16 slaves)} \\
     \midrule
     \textbf{Item} & avg & std & max & avg & std & max & avg & std & max & avg & std & max \\
     \hline
     OpenPLC & 2.9 & 0.2 & 5.0 & 5.5 & 0.2 & 7.8 & 10.8 & 0.4 & 16.7 & 21.2 & 0.7 & 30.3 \\
     \hline
     Enhanced TEE-PLC & 9.5 & 0.2 & 10.2 & 17.3 & 0.3 & 19.8 & 32.7 & 0.6 & 37.6 & 63.6 & 0.8 & 73.9 \\
     \hline
     Minimal TEE-PLC & 2.9 & 0.2 & 4.5 & 6.6 & 0.3 & 8.9 & 11.0 & 0.5 & 16.8 & 21.6 & 0.7 & 31.7 \\

     \bottomrule
    \end{tabular}
\end{table*}





\subsection{Performance and Overhead}

To get a precise timestamp for measurement, we employed different functions in both the normal world and the secure world. In the normal world, a Linux system call {\it clock\_gettime()} can obtain a timestamp with nanosecond-level precision and introduces only 200ns latency on Raspberry Pi. 
In the secure world, GlobalPlatform API {\it TEE\_GetSystemTimer()} can only offer millisecond granularity, so we employed a physical counter of ARM CPU ({\it CNTPCT\_EL0}) to get high-res timestamps. This physical counter runs at a frequency of 19.2MHz and only introduces a latency of 40ns. 

We ran the same control logic first on OpenPLC and then on both Enhanced and Minimal TEE-PLC 1000 times and calculated the average duration per scan cycle. The measurements with different numbers of slave devices are shown in Table~\ref{table:exec_time}. 
The average overhead owing to the use of TEE is about 6ms for every two connected slave devices (a pair of sensor and actuator). 
As seen in Table~\ref{table:exec_time}, the latency of the scan cycle on Enhanced TEE-PLC grows linearly to the number of slave devices. 
To evaluate practicality, we studied real-world PLC logic in a filling tank system~\cite{attkfinder} and the generator synchronization in a power grid~\cite{falsedata}. {\color{black} Detailed description of these systems are provided in Appendix B.} Such PLCs interact with up to 4 slaves. The overall latency for each scan cycle of the Enhanced TEE-PLC is less than 20ms. According to ~\cite{plccycle}, the scan cycle time for safety PLCs is 20ms, and one for process control PLC is a few hundred milliseconds. Therefore, Enhanced TEE-PLC can still meet the real-world safety requirements of this scale and can support a larger number of slave devices for the process control.
{\color{black}Discussions in the specific ICS contexts considered are also found in Appendix B.} 
%
The Minimal TEE-PLC does not incur overhead on the communication and shows performance similar to the original OpenPLC.
\begin{figure}[!t]
    \centering
    \vspace{-4mm}
    \includegraphics[width=0.48\textwidth]{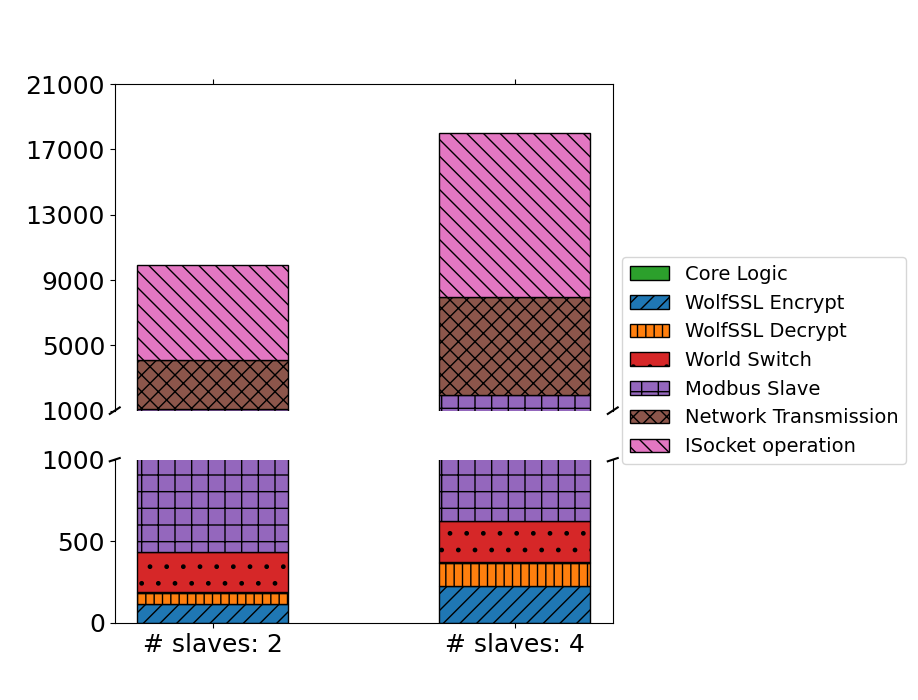}
    \vspace{-8mm}
    \caption{Enhanced TEE-PLC Latency Breakdown ($us$)}
    \Description{}
    \label{fig:latency_contrib}
        \vspace{-5mm}

\end{figure}



We further studied the detailed breakdown of the scan cycle latency of the Enhanced TEE-PLC. The results with 1 pair and 2 pairs of slaves are shown in Figure~\ref{fig:latency_contrib}. The network transmission latency, as well as the processing time of Modbus slaves, is measured from Wireshark traces.
The average latency of a single round-trip world switch is 280us for both cases. As expected, latency for TLS processing ({\it WolfSSL\_Encrypt} and {\it WolfSSL\_Decrypt}), network transmission latency and Modbus slave processing time increase linearly to the number of slaves. We also see that the duration of control logic execution itself is negligible (around 5us). 
One notable finding is that the latency for handling communication using TEE\_iSocket is significant, as discussed further in Section~\ref{sec:disc}.

\subsection{Binary Size and Memory Usage}
Both Minimal and Enhanced TEE-PLC consist of the modified OpenPLC runtime in the normal world and the Logic-only TA or Scan-cycle TA in the secure world, respectively. Compared with the original OpenPLC taking 206.3 KiB as a single file on disk, the Enhanced TEE-PLC is 586.5 KiB in total, composed of 113.2 KiB for modified OpenPLC runtime and 473.3 KiB for Scan-cycle TA. OP-TEE itself also introduces an extra 402.5 KiB storage footprint to the Linux kernel. In the normal world file system, OP-TEE consumes an extra 390.4 KiB, including the client application APIs (such as {\it libteec}, {\it libckteec}, {\it libseteec}) and a Linux user space supplicant daemon TEE-supplicant. 
On the other hand, the size of Logic-only TA is 62.8KiB, which demonstrates the smaller TCB. 

The runtime memory consumption of the Enhanced TEE-PLC, measured with the $top$ command, is shown in Table~\ref{table:mem_usage}. Compared with the original OpenPLC, which consumes 80.2 MiB RAM only in the normal world, the Enhanced TEE-PLC needs 92.5 MiB memory in total: 14.5 MiB for OpenPLC runtime and 76.4 MiB for TEE-supplicant in the normal world as well as 803 KiB for Scan-cycle TA and 857 KiB for OP-TEE OS in the secure world. The Minimal TEE-PLC, which not only shares the same Modbus/TCP communication module as the original OpenPLC but also relies on TEE-supplicant for invoking secure services from the normal world, consumes the most memory among all listed implementations.
%
\begin{table}
  \caption{Memory Usage Breakdown (MiB)}
\vspace{-2mm}
  \label{table:mem_usage}
  \begin{tabular}{||c|c|c|c||}
    \toprule
    \textbf{Component} & \textbf{OpenPLC} & \textbf{Enhanced} & \textbf{Minimal} \\
    \hline
    PLC Runtime & 80.2 & 14.5 & 80.2 \\
    \hline
    TEE-supplicant & NA & 76.4 & 76.4 \\
    \hline
    TA & NA & 0.8 & 0.1 \\
    \hline
    OP-TEE OS & NA & 0.8 & 0.8 \\
    \hline
    Total & 80.2 & 92.5 & 157.5 \\
    \bottomrule
  \end{tabular}
\vspace{-3mm}
\end{table}
%

\section{Discussion and Recommendations}
\label{sec:disc}


\subsection{Directions for Performance Improvement}

Based on our measurements shown in Section~\ref{sec:eval}, the major portion of the overhead is attributed to iSocket operation. According to OP-TEE specification~\cite{op-tee}, at the high level, the processing of network packets sent by the TA involves iSocket Pseudo TA in the secure world and TEE-supplicant in the normal world. 
Based on our study, iSocket Pseudo TA simply forwards packets from/to TEE-supplicant, and thus its latency is negligible.

Moreover, for both of the Enhanced TEE-PLC and the Minimal TEE-PLC, TEE-supplicant consumes a non-negligible amount of memory and the amount is the same for both implementations. 
Based on our observation, immediately after any TA is loaded, it starts to consume this amount of memory (regardless of whether communication with the slaves is involved or not).
By implementing the Scan-cycle TA as ``early TA''~\cite{op-tee}, which is linked to the TEE core and thus does not involve TEE-supplicant for invocation, we could eliminate the reliance on it when loading the TA. In the case of the Minimal TEE-PLC, which does not rely on the TEE-supplicant for network communication, the memory usage by TEE-supplicant can be reduced. 
One drawback of early TA is that update of the Logic-only TA (i.e., PLC logic) becomes difficult after device deployment.
%
%

TEE-supplicant is outside of the GlobalPlatform specification (i.e., OP-TEE specific), and thus the use of other secure OS implementations could help. When vendors design a secure OS for PLCs, optimization of this part is strongly recommended.

Another observation is that the communication latency increases linearly to the number of slave devices for the Enhanced TEE-PLC. 
The bottleneck here is that a TA {\color{black}on OP-TEE cannot be multi-threaded~\cite{wan2020rustee}}, and thus, communication sessions must be executed sequentially. 
One potential improvement for supporting more slave devices is to separate the secure communication from the Scan-cycle TA and make it a standalone multi-instance TA. Multi-instance TAs run on different cores in parallel and could be invoked by multi-threaded normal world applications concurrently and thus can reduce the overall latency.
However, the parallelism of multi-instance TA is limited to the number of cores of embedded SoCs. Since single or dual-core SoC is still the mainstream of PLCs, the improvement brought by multi-instance TAs may not be significant.

\subsection{Reliance on Normal World Components}

Isolation is the main philosophy of TEE security design. TrustZone and OP-TEE protect sensitive data and code by implementing both spatial and temporal isolation of the secure world from the normal world. However, some fundamental functions of TEE still rely on normal world components, such as loading TA files from the normal world file system, shared memory allocation, and network connection. 
In OP-TEE, TEE-supplicant is an indispensable component running in the normal world to perform such tasks.
Hence, TEE-supplicant may be the weak point of the system. 
 For instance, attackers may kill the daemon and lead to no response from the whole secure world. 
{\color{black}A potential solution to address this threat is to use a secure timer feature provided by ARM TrustZone. In ARM TrustZone-aware SoCs, physical timers can be configured as secure timers whose configuration cannot be modified from the normal world. 
The periodical secure timer interrupts ensure the secure world to regain control of the CPU at a specific time. By doing this, the trusted applications no longer rely on their normal-world partner for invocation but are awakened by secure timer interrupts to ensure a periodical execution of control logic~\cite{busch2019teemo}. This design counter one kind of DoS attack. However, this alone is not a complete solution. Compromised TEE-supplicant may block the network communication from the Scan-cycle TA to actuators which implies another kind of DoS attack. Thus, it is recommended to deploy an external network monitor or intrusion detection system to check the periodicity of communication initiated by the PLC device as well as the occurrence of network error.}

\subsection{Securing PLC's Server Interface}
In the Enhanced TEE-PLC architecture, there is a remaining security risk. PLCs usually send data collected from sensors in the field to SCADA HMI. For such communication, a PLC implements a server interface
so that other devices can query data. However, according to GlobalPlatform's specification on TEE\_iSocket~\cite{gp2021teesocket}, the server socket is not supported in the secure world. This means that the server interface for other devices needs to rely on the normal world,
%
which may be under attacker's control.
We admit that this is the limitation of the current implementation. Because of this limitation, while the integrity of control logic execution is ensured, the data sent to the SCADA HMI could be subject to malicious manipulation in the normal world.

{\color{black}\subsection{Relationship to RT-TEE~\cite{RT-TEE}}}
{\color{black}
The lack of a scheduler in OP-TEE OS means there is no guarantee of real-time execution for secure tasks.
RT-TEE~\cite{RT-TEE} is built on top of OP-TEE and implements a two-layer policy-based event-driven hierarchical scheduler to enforce real-time scheduling and an I/O reference monitor for the availability of hardware/software interactions. The authors also proposed two different approaches to port drivers into RT-TEE by debloating trusted drivers and sandboxing untrusted drivers. Since RT-TEE retains the same interfaces for OP-TEE TAs, Both TEE-PLCs can also benefit from real-time scheduling and hardened I/O security from RT-TEE. In addition, if the network driver can be implemented as a sandboxed untrusted driver, the Enhanced TEE-PLC would not need to rely on iSocket and thus would attain a better performance. 

However, to take full advantage of RT-TEE, secondary developers must conduct comprehensive investigations of their use cases to create both real-time scheduling and I/O access control policies. Moreover, porting drivers to RT-TEE is not always trivial. Although debloated trusted drivers are basically templates created from historical interaction data between hardware and drivers and are thus easy to create, it assumes that the interaction between hardware and software is simple and highly predictable. Templates of complicated scenarios like error handling and data encryption are definitely hard to create and error-prone. As for sandboxed untrusted drivers, secondary developers need to manually split the driver into secure and non-secure halves and instrument the secure part to enable sandboxing. 
{\color{black} Porting of drivers is not a trivial task for many users in research community without supports from vendors. In order to provide an accessible platform for research and evaluation, we chose OP-TEE for our open-source implementation, instead of RT-TEE.}
In the TEE-PLC implementations,
we left the network drivers totally in normal world and thus avoided such difficulty. Moreover, feasibility of implementing realistic PLC functionality with support of standardized ICS network protocols on top of RT-TEE should be validated though prototyping. 

\color{black}
\subsection{Real-world Implications}
Typically, a real PLC consists of CPU + memory and input/output modules. Among them, our TEE-PLC design and implementation using Linux, OpenPLC, and OP-TEE fits ``CPU + memory'' part, which is responsible for operating on field devices connected via I/O modules and also hosts control logic. While actual codes need to be tailored to each PLC platform by a PLC vendor,
the overall idea (e.g., separation of secure and normal world modules and interface between them) can be translated from our implementation using OP-TEE and OpenPLC. 
 Moreover, the performance presented in Section~\ref{sec:eval} can be seen as the lower bound of the TrustZone-based PLC design discussed in this paper, since the real PLC products to appear in the market may be able to potentially benefit from real-time OS (e.g., FreeRTOS~\cite{freertos}) and RT-TEE. Given that our performance measurement already shows practicality for small, but realistic, ICS settings, the use of ARM TrustZone for securing PLC integrity is deemed promising.

}


\section{Related Work}\label{sec:relatedwork}

{\color{black}Use of ARM TrustZone TEE for PLC is not completely a new idea. For instance, Denzel et al.~\cite{denzel2017malware} utilized FreeRTOS, the ARM TrustZone-aware real-time OS~\cite{freertos}, for PLCs. Their purpose is auto-healing of the system, and they utilize the secure world to store the PLC software image for automated restoration. However, PLC scan cycle is still executed in the normal world, which is subject to attacks in our scope. Thus, it is orthogonal to our contribution.
}






An intrusion detection system (IDS) is a monitoring system that detects suspicious activities in the network. In the literature, we can find a number of efforts of IDSes for ICS.  
For instance, \cite{tan2022cotoru} proposes using PLC logic to generate intrusion detection rules to be implemented on network-based IDSes.
Bohara et al.~\cite{bohara2020ed4gap} utilized information in the IEC 61850 GOOSE message payload and detected anomalies using state machines. Another protocol-specific effort can be found in~\cite{lin2013adapting}. Multi-model anomaly detection, which combines protocol-specific information as well as statistics derived from network traces, is proposed in~\cite{ren2018edmand}. 
One major limitation of these network-based IDSes is that it is not capable of detecting attackers on ICS devices (e.g., malicious firmware, rootkit, etc.) who can compromise the integrity of the control logic execution as well as steal information about the control logic. 
Even though the host-based IDSes could detect such attacks, the limited computational resources on PLCs would prohibit deployment of them. 
%
%
%
Moreover, as pointed out by Khraisat et al.~\cite{khraisat2019survey}, 
IDSes cannot work well with encrypted traffic (e.g., TLS/SSL) since the encryption impedes detailed inspection of packet payloads. 


In order to counter the injection or manipulation of ICS message payloads, message authentication solutions for ICS have been proposed. 
%
For example, in power grid sectors, IEC Technical Committee developed the data and communication security standard IEC 62351 to provide end-to-end communication security for smart grid communications. Among the standards, IEC 62351-3 recommends the usage of TLS 
to provide protection for protocols over TCP/IP, including DNP3, Modbus TCP, IEC 60870-5-104, IEC 61850 MMS, and so forth.
The academic and research community also proposed a number of solutions for low-latency message authentication for various communication models in ICS~\cite{tefek2022caching,esiner2019f,10.1145/3607194,esiner2022lomos}.
However, end-to-end communication with encryption and authentication mechanisms is not sufficient to counter our attacker models. During the process of encryption and decryption of TLS, the payload in memory can be tampered with by attackers on the device.
Therefore, the endpoint of the secure communication should be moved into the secure world and done in the secure memory space. In addition, private keys and other sensitive data used for secure communication stored on the device storage are also at risk.
Our design allows such sensitive data to be stored in secure storage
so that the sensitive data will not be leaked. 

Control Flow Integrity (CFI) is an effective defense against control flow hijacking. CFI ensures that a program only follows its control-flow graph (CFG) as approved execution paths which is determined during compilation time. This can be achieved by checking the destination address of each indirect branch/call and return instructions. The control logic injection is essentially one kind of control flow hijacking and can be solved by CFI theoretically. However, most of the CFI techniques are designed for general-purpose computers and are not suitable for embedded devices. While some research tailored CFI for embedded devices~\cite{davi2012mocfi,pewny2013control}, exception handlers and loops are used for CFG verification and thus introduce considerable overhead. To ensure the availability and real-time requirement of PLC, Abbasi et al.~\cite{abbasi2017ecfi} proposed an asynchronous and non-blocking CFI that works well for resource-constraint devices. However, similar to other processes running on PLC, the integrity of CFI itself can be compromised by attackers with root access, and thus the protection for PLC control logic can lose effectiveness.

Remote attestation is a technology for enabling a verifier device to check the integrity of software (or firmware) on a device of interest (i.e., prover); efforts are made in the context of ICS, e.g.,~\cite{chen2017secure}. Using remote attestation for PLCs, the malicious modification of the software can be detected, but it does not necessarily prevent the modification of the logic. Moreover, because of the overhead for the verification task on a prover ICS device, typically, the attestation is often done infrequently, which could lead to a TOCTOU (time-of-check to time-of-use) problem. 

\section{conclusions}
\label{sec:concl}

Programmable logic controllers (PLC) are universally found in industrial control systems (ICS) nowadays for implementing closed control loops. Because of the crucial roles that PLCs play, they often become targets of attacks against critical infrastructure. While several security technologies have been proposed, such solutions were not effective against attackers who have root access to the devices.
In this paper, we discussed attack vectors against the integrity of the control loop implemented by a PLC and discussed mitigation using ARM TrustZone TEE. Furthermore, through the proof-of-concept design and implementation of a PLC hardened by TEE, named TEE-PLC, {\color{black} which is open-sourced for further study and evaluation,} we demonstrated the feasibility and practicality under realistic ICS settings. 
Through the study based on the OP-TEE secure OS, we also discussed desired properties in the secure OS design 
for 
latency-stringent operation in ICS of a larger scale.
%
Our proof of concept focused on the security of 
PLC's scan cycle, and thus other features needed on a PLC product are still missing.
Having this said, we hope this study provides directions towards the practical usage of ARM TrustZone TEE for full-fledged PLCs and other ICS devices.

\begin{acks}
This research is supported by the National Research Foundation, Prime Minister’s Office, Singapore under its Campus for Research Excellence and Technological Enterprise (CREATE) programme. 
The authors would also like to thank the shepherd and anonymous reviewers for their  insightful, constructive comments for improving the technical quality and presentation of this paper.
\end{acks}

\bibliographystyle{ACM-Reference-Format}
\bibliography{reference}


\appendix
\section*{APPENDIX A: Process Flow of TEE-PLC}
{\color{black}We here elaborate the implementation details of TEE-PLC. We focus on discussion on Enhanced TEE-PLC because it is more complicated, and then briefly discuss Minimal version.}

\begin{figure}[!h]
  \includegraphics[width=\linewidth]{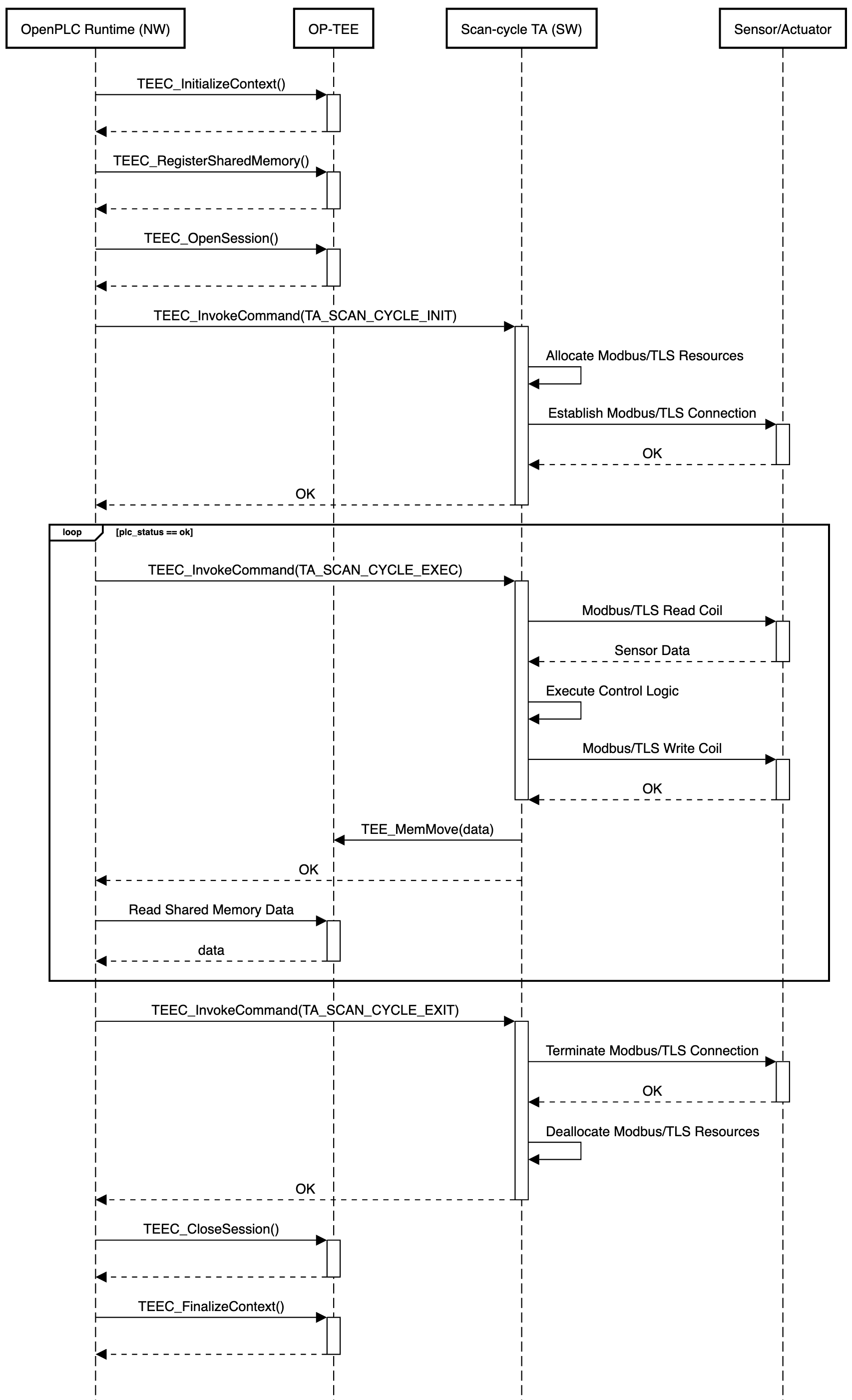}
  \caption{Enhanced TEE-PLC Process Flow}
  \label{fig:data_flow}
\end{figure}

In Enhanced TEE-PLC, the process flow of Scan-cycle TA  and its interaction with the OpenPLC runtime in the normal world is depicted in Figure~\ref{fig:data_flow}. For simplicity, we use the generic OP-TEE to refer to OP-TEE OS, TEE-supplicant, and OP-TEE driver and omit the underlying jobs of OP-TEE. Meanwhile, only the ideal case is considered, and error handling is not shown.

The entry points we defined in the Scan-cycle TA include, {\it TA\_SCAN \_CYCLE\_INIT} for resource allocation as well as Modbus/TLS connection, {\it TA\_SCAN\_CYCLE\_EXEC} for scan cycle execution, and {\it TA\_SCAN\_CYCLE\_EXIT} for Modbus/TLS disconnection and resource recycling. 
When TEE-PLC starts to run, OpenPLC runtime in the normal world first notifies OP-TEE to set up the TEE context and initialize the TA session for obtaining services from the Scan-cycle TA. During this stage, OP-TEE loads the Scan-cycle TA file from REE-FS (rich execution environment file system) to secure memory decrypts and checks the signature of the TA file before running it. In the meanwhile, OpenPLC runtime also registers shared memory for gathering information of scan cycles running in the secure world. This helps engineers to monitor whether the underlying industrial processes are running correctly in the secure world. 

Once the prerequisites are all met, OpenPLC runtime invokes the entry point of {\it TA\_SCAN\_CYCLE\_INIT}. As a response, the Scan-cycle TA establishes Modbus/TLS connection and allocates resources needed by subsequent scan cycles. If no error returns, OpenPLC runtime periodically launches the scan cycle and invokes the entry point of {\it TA\_SCAN\_CYCLE\_EXEC}. Once receiving the request, the Scan-cycle TA sequentially queries each connected sensor through Modbus/TLS communication and collects data used by control logic. It then executes control logic and sends out commands to its connected actuators through {\it Modbus\_TLS\_WriteCoil()}. When this round of scan cycle is about to finish, TA copies the execution outcomes to shared memory through {\it TEE\_MemMove()}. 

In our design, the Scan-cycle TA session keeps alive after it is established. Modbus/TLS connections between the TA and sensors and actuators also never shut down for each scan cycle execution. This design significantly reduces the overhead caused by repeated TLS handshakes. When an internal error happens, or maintenance of the TA is required, OpenPLC runtime can shut down the scan cycle through {\it TA\_SCAN\_CYCLE\_EXIT}. The Scan-cycle TA shuts down the Modbus/TLS connection and deallocates the resources for the previous scan cycles.
Finally, OpenPLC runtime in the normal world notifies OP-TEE to close the TA session and finalize the TEE context.

{\color{black}
The implementation of Logic-only TA in Minimal TEE-PLC is much simpler. Logic-only TA implements only one entry point, {\it TA\_CONTROL\_LOGIC}, which executes the PLC control logic. Communication with external sensors and actuators are handled by the OpenPLC runtime in the normal world, and for each scan cycle, the OpenPLC runtime collects data from sensors, and then invokes Logic-only TA, passing collected sensor data via the shared memory. The outcome of the logic execution in the secure world is returned to the normal world via the shared memory, and finally the OpenPLC runtime sends out control commands to actuators when necessary. 
}

{\color{black}
\section*{Appendix B: Practical ICS Contexts Considered for Evaluation}
In order to evaluate if the measured latency is still in a practical range, we consider some real-world PLC control logic examples.

\begin{figure}[h]
    \centering
        \vspace{-4mm}

    \includegraphics[width=\linewidth]{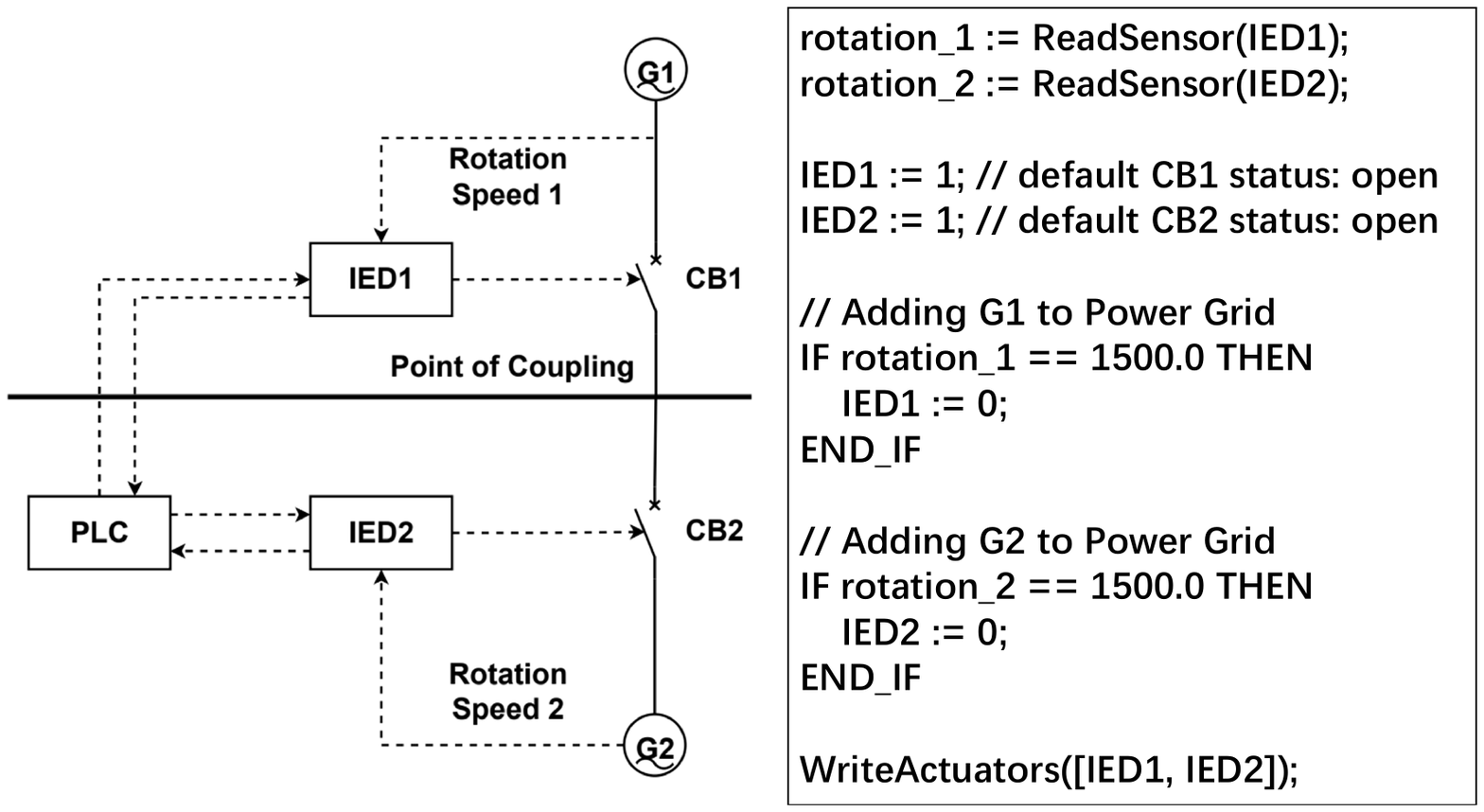}
    \vspace{-10mm}
    \caption{PLC Logic for Generator Synchronization in Power Grid System~\cite{falsedata}}
    \Description{}
    \label{fig:gen_sync}
\end{figure}

\begin{figure}[h]
    \centering
    \vspace{-10mm}

    \includegraphics[width=\linewidth]{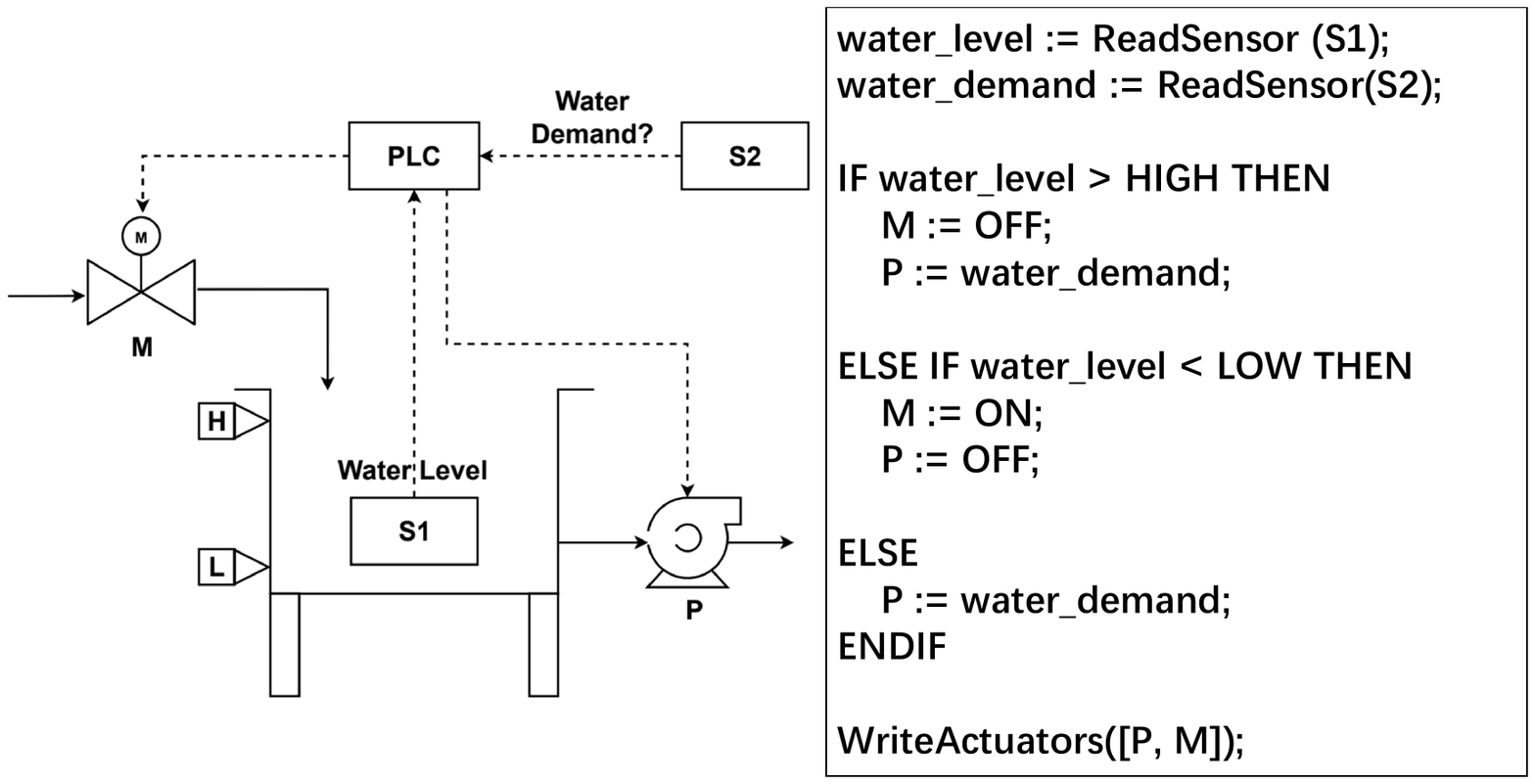}
    \vspace{-12mm}
    \caption{PLC Logic for Filling Tank System~\cite{attkfinder}}
    \Description{}
    \label{fig:fill_tank}
\end{figure}

\noindent \textbf{Generator Synchronization  in Power Grid~\cite{falsedata}:}
In an AC power grid, synchronization is the process of matching the rotation speed 
between the newly-started generator and the grid already in operation. The example of the synchronization process in EPIC testbed~\cite{epic,falsedata} and its corresponding PLC logic is shown in Figure~\ref{fig:gen_sync}. 
In the initial state, the circuit breakers (CB1 and CB2) are open and the new generators are disconnected from the power grid. Intelligent Electronic Devices (IED1 and IED2) measure
the rotation speed of the two generators and report them to the PLC.
When the difference is less than a threshold, the PLC instructs the IEDs to close the CBs to connect the new generators to the grid. 
In this example, the PLC is connected to 2 IEDs. Since each IED works as a sensor (monitoring phase angle) and actuator (opening/closing CB), we regard these 2 IEDs as 4 connected devices. Therefore, the power grid synchronization process can be carried out by TEE-PLC at a rate of 50Hz. We should note that, very time-sensitive protection and control, which requires millisecond-level latency limit (e.g., opening circuit breakers under over-current/voltage situations)~\cite{ieee1646} are typically done by IEDs, and control implemented by PLCs, including generator synchronization discussed here, is less stringent. 

\noindent \textbf{Filling Tank System~\cite{attkfinder}:} This is a water supply system to drain water on users' demands while maintaining the water level within a specific range, as shown in Figure~\ref{fig:fill_tank}. PLC queries 
sensor S2
whether there is demand for water and turns on pump P if the reply is True. In the meantime, it will open motor M if the current water level reading from S1 is low. In this system, the PLC interacts with 2 sensors and 2 actuators. According to Table~\ref{table:exec_time}, the average latency for a scan cycle is less than 20ms, which is the default execution interval of OpenPLC. 
Moreover, a water plant is in general less latency stringent. For instance, as discussed in the attack experiments using SWaT testbed, a state-of-the-art water treatment system testbed~\cite{mathur2016swat}, the timescale is in the order of seconds~\cite{urbina2016attacking}. Thus, 20ms overhead is considered small enough in practice.






}

\end{document}